\documentclass[conference,a4paper]{IEEEtran}
\usepackage{enumerate}
\usepackage[top=0.8in, bottom=1.5in, left=0.6in, right=0.6in]{geometry}
\usepackage{epstopdf}
\usepackage{cite}
\usepackage{amscd}
\usepackage{dsfont}
\usepackage{latexsym}
\usepackage{array}

\usepackage{tikz}\usetikzlibrary{calc}
\usepackage{tabularx}
\usepackage{epsfig}
\usepackage{graphicx,subfigure}
\usepackage{amsmath,mathrsfs}
\usepackage{amsthm}
\usepackage{amssymb}  
\usepackage{amsbsy}
\usepackage{color}
\usepackage{stfloats}
\usepackage{algorithm}
\usepackage{algpseudocode}
\usepackage{bm}
\usepackage{pifont}
\usepackage{accents}
\usepackage{multirow}

\newtheorem{lem}{Lemma}
\newtheorem{ther}{Theorem}
\newtheorem{deft}{Definition}
\newtheorem{prop}{Proposition}

\theoremstyle{definition}
\newtheorem{rem}{Remark}

\ifCLASSINFOpdf
\else
\fi
\begin{document}
%
\title{On the Quantization Goodness of Polar Lattices}

\author{
\IEEEauthorblockN{Ling Liu}
\IEEEauthorblockA{Xidian University\\
Xi'an, China\\
Email: liuling@xidian.edu.cn}
\and
\IEEEauthorblockN{Shanxiang Lyu}
\IEEEauthorblockA{Jinan University\\
Guangzhou, China\\
Email: lsx07@jnu.edu.cn}
\and
\IEEEauthorblockN{Cong Ling}
\IEEEauthorblockA{Imperial College London\\
London, UK\\
Email: cling@ieee.org}
\and
\IEEEauthorblockN{Baoming Bai}
\IEEEauthorblockA{Xidian University\\
Xi'an, China\\
Email: bmbai@mail.xidian.edu.cn}
}


%


\maketitle

\begin{abstract}
In this work, we prove that polar lattices, when tailored for lossy compression, are quantization-good in the sense that their normalized second moments approach $\frac{1}{2\pi e}$ as the dimension of lattices increases. It has been predicted by Zamir et al. \cite{ZamirQZ96} that the Entropy Coded Dithered Quantization (ECDQ) system using quantization-good lattices can achieve the rate-distortion bound of i.i.d. Gaussian sources. In our previous work \cite{LingQZ}, we established that polar lattices are indeed capable of attaining the same objective. It is reasonable to conjecture that polar lattices also demonstrate quantization goodness in the context of lossy compression. This study confirms this hypothesis.
\end{abstract}


%
\IEEEpeerreviewmaketitle

\section{Introduction}
Modern digital communication systems heavily rely on analog-to-digital (AD) converters to process raw analog signals at their interface. The quest for good AD converter or source quantizer may trace its roots back to Lloyd's pioneering work in \cite{lloyd}, which introduced an iterative algorithm for the optimal design of fixed-rate scalar quantizer. Then it was extended to higher dimensions, that is, vector quantization (VQ)\cite{LindeVecQ1980,GrayVecQz1984}. The rationale behind the pursuit of higher-dimensional quantization stemmed from Shannon's rate-distortion theory \cite{ShannonRD1959}, which characterizes the optimal balance between compression rate and quantization-induced distortion. Many practical VQ techniques have emerged over the past few decades. Lattice VQ, in particular, has garnered significant attention and was systematically studied in \cite{VoronoiQzC&S1984}, where the authors built a bridge between the second moments of lattices and the performance of source coding. The appeal of lattice VQ may attribute to its highly regular structure, facilitating compact storage and swift implementation.

The Entropy Coded Dithered Quantization (ECDQ) system \cite{Ziv1985} consists of a lattice quantizer, an entropy encoder and its matched entropy decoder, among which a random dither is shared. The sharing dither helps to remove certain effect of the source statistics, leading to several universal quantization schemes. In particular, for the i.i.d. Gaussian sources, when combined with appropriate pre/post filtering, the ECDQ system was shown to be able to achieve the rate-distortion bound under the quadratic distortion if the underlying lattices are quantization-good, i.e., their Normalized Second Moments (NSMs) (defined in Sect. \ref{sec:background}) approach $\frac{1}{2\pi e}$. The NSMs of some classical lattices were calculated and reported in \cite{yellowbook}. Some more recent results in this direction can also be found in \cite{ErikBruceQz23,ShanxiangQZ22,ErikBruceBWL23}. In the context of lattice coding over Additive White Gaussian Noise (AWGN) channels, it has also been proved that using quantization-good lattices for shaping can obtain the optimum shaping gain \cite{ZamirAchieve1/2}. Consequently, the explicit construction of quantization-good lattices has become a challenge being paid closed attention to in recent years.

In \cite{BK:Rogers}, Rogers defined the lattices that are good for covering (referred to as ``Rogers good'') and proved their existence. The existence of quantization-good lattices was guaranteed, as covering-goodness indeed implies quantization-goodness \cite[Corollary 7.3.1]{BK:Zamir}. A more direct proof of their existence is to use random ensembles of construction A lattices \cite{zamir2}, where lattices are constructed based on $q$-ary random linear codes. An improvement was recently given in \cite{LDAgood17}, with random ensembles of construction A lattices restricted to random ensembles of Low Density construction A (LDA) lattices. More explicitly, the authors claimed that under certain hypotheses (See \cite[Thm. 3]{LDAgood17}.), a sequence of lattices chosen from the random ensembles of LDA lattices is quantization-good with probability 1. In this work, we propose a more deterministic structure over the lattices, i.e. polar lattices \cite{yan2}, and prove that a sequence of polar lattices can be quantization-good as the lattice dimension increases, without extra hypotheses.

As an explicit counterpart of the good random lattice ensembles, polar lattices have been constructed to achieve the rate-distortion bound of i.i.d. Gaussian sources \cite{LingQZ}. Unlike the construction A lattices, a polar lattice is generated from a set of nested polar codes \cite{arikan2009channel} and a lattice partition chain, following the construction D method \cite{yellowbook}. This method allows us to use binary linear codes to construct high dimensional lattices from significantly lower dimensional lattice partition chains. Using the lattice Gaussian distribution as the reconstruction distribution \cite{LingBel13}, the sharing dither in the ECDQ system can be removed, as has been shown in \cite{CongQZ}. Moreover, thanks to the source polarization technique \cite{polarsource,cronie2010lossless}, the entropy encoding process can be integrated into the quantization process of polar lattices. To be compatible with the multilevel structure of construction D lattices, the entropy encoding of the reconstruction points is performed according to their binary labeling along the partition chain level by level, which can be viewed as the shaping operation for polar lattices. As a result, the shaped polar lattice quantizer was proved to be rate-distortion bound achieving for i.i.d. Gaussian sources, while the properties of the underlying polar lattice (without shaping) was not explicitly examined. The goal of this work is to fill this gap, and to demonstrate that polar lattices are quantization-good in the context of lossy compression.

The rest of the paper is organized as follows. Sect. II gives preliminaries of polar codes, lattices and discrete Gaussian distribution. The polar lattice quantizer for i.i.d. Gaussian sources is revisited in Sect. III, where we leach out the lattice shaping operation integrated in the previous work \cite{LingQZ}. This allows for a deeper investigation into the characteristics of the lattice itself. In Sect. IV, we present our main argument on the quantization goodness of polar lattices. The paper is concluded in Sect. V.

$\it{Notation:}$ All random variables (RVs) are denoted by capital letters. Let $P_X$ denote the probability mass function of a RV $X$ taking values in a countable set $\mathcal{X}$, and the probability density function of a RV $Y$ in an uncountable set $\mathcal{Y}$ is denoted by $f_Y$. The combination of $N$ i.i.d. copies of $X$ is denoted by a vector $X^{1:N}$ or $X^{[N]}$, where $[N]=\{1,...,N\}$, and its $i$-th element is given by $X^i$. The subvector of $X^{[N]}$ with indices limited to a subset $\mathcal{F} \subseteq [N]$ is denoted by $X^{\mathcal{F}}$. The cardinality of $\mathcal{F}$ is $|\mathcal{F}|$. A lattice partition chain is denoted by $\Lambda (\Lambda_0)/\Lambda_1/\cdots/\Lambda_i/\cdots$, where $\Lambda_i$ is an $n_p$-dimensional lattice. For multilevel polar coding and polar lattices, we denote by $X_{\ell}$ a binary representation random variable of $X$ at level $\ell$. The $i$-th realization of $X_{\ell}$ is denoted by $x_{\ell}^{i}$. We also use the notation $x_{\ell}^{i:j}$ as a shorthand for a vector $(x_{\ell}^{i},...,x_{\ell}^{j})$, which is a realization of random variables $X_{\ell}^{i:j}=(X_{\ell}^{i},...,X_{\ell}^{j})$. Similarly, $X_{\ell:\jmath}^{i}=(X_{\ell}^{i},...,X_{\jmath}^{i})$. We use $\log$ for binary logarithm throughout this paper.

\section{Preliminaries of Polar Lattices}\label{sec:background}

\subsection{Polar Codes}
Let $\tilde{W}$ be a binary-input memoryless symmetric channel (BMSC) with input $X \in \mathcal{X}$ and output $Y \in \mathcal{Y}$. Given the capacity $C$ of $\tilde{W}$ and a rate $R<C$, the information bits of a polar code with block length $N=2^m$ are indexed by a set of $\lfloor RN \rfloor$ rows of the generator matrix $G_N=\left[\begin{smallmatrix}1&0\\1&1\end{smallmatrix}\right]^{\otimes m}$, where $\otimes$ denotes the Kronecker product. The matrix $G_N$ combines $N$ identical copies of $\tilde{W}$ to $\tilde{W}_N$, which can be successively split into $N$ binary memoryless symmetric subchannels, denoted by $\tilde{W}_{N}^{(i)}$ with $1 \leq i \leq N$. By channel polarization \cite{arikan2009channel}, the fraction of good (roughly error-free) subchannels approaches $C$ as $m\rightarrow \infty$. Thus, to achieve the capacity, information bits are transmitted over those good subchannels, while the rest are assigned with frozen bits known at the receiver before transmission. The indices of good subchannels can be identified based on their associated Bhattacharyya parameters.
\begin{deft}\label{deft:symZ&asymZ}
Given a BMSC $\tilde{W}$ with transition probability $P_{Y|X}$, the Bhattacharyya parameter $\tilde{Z}$ of $\tilde{W}$ is defined as
\begin{eqnarray}
\tilde{Z}(\tilde{W})=\tilde{Z}(X|Y)&\triangleq\sum\limits_{y} \sqrt{P_{Y|X}(y|0)P_{Y|X}(y|1)}.
\end{eqnarray}
When $W$ is a binary-input memoryless asymmetric channel with input distribution $P_X$, the above definition is generalized as
\begin{eqnarray}
Z(W)=Z(X|Y)&\triangleq 2\sum\limits_{y} \sqrt{P_{X,Y}(0,y)P_{X,Y}(1,y)}.
\end{eqnarray}
\end{deft}

We note that both $\tilde{Z}$ and $Z$ lie within the range $[0,1]$, and $Z(X|Y) = \tilde{Z}(X|Y)$ holds when $P_X$ is unbiased.

In \cite{KoradaSource}, when constructing polar codes for the lossy compression of a binary symmetric source, the information set $\tilde{\mathcal{I}}$ is chosen as $\{i\in [N]:\tilde{Z}(\tilde{W}_{N}^{(i)}) < 1-2^{-N^{\beta}}\}$ for a given constant $0<\beta <\frac{1}{2}$, and the frozen set $\tilde{\mathcal{F}}=\tilde{\mathcal{I}}^c$ is the complement of $\tilde{\mathcal{I}}$. Here, $\tilde{W}$ is the test channel corresponding to a target distortion, and $\beta$ is commonly called the rate of polarization \cite{arikan2009rate} for the 2-by-2 kernel $\left[\begin{smallmatrix}1&0\\1&1\end{smallmatrix}\right]$. We note that the Bhattacharyya parameter is still written as a form of summation for continuous $Y$, because in practice a certain degree of discretization over $Y$ is needed, especially for large block length. Efficient algorithms to evaluate the Bhattacharyya parameter of subchannels for general BMSCs were presented in \cite{tal2011construct,PolarConstru,mori2009performance}.

\subsection{Lattice Codes and Polar Lattices}
An $n$-dimensional lattice is a discrete subgroup of $\mathbb{R}^{n}$ which can be described by
\begin{eqnarray}
\Lambda=\{ \lambda=\mathbf{B}\mathbf{z}:\mathbf{z}\in\mathbb{Z}^{n}\},
\end{eqnarray}
where the columns of the generator matrix $\mathbf{B}=[\mathrm{b}_{1}, \cdots, \mathrm{b}_{n}]$ are assumed to be linearly independent.

For a vector ${x}\in\mathbb{R}^{n}$, the nearest-neighbor quantizer associated with $\Lambda$ is $Q_{\Lambda}({x})=\text{arg}\min\limits_{ \lambda\in\Lambda}\|\lambda-{x}\|$. The Voronoi region of $\Lambda$ around 0, defined by $\mathcal{V}(\Lambda)=\{{x}:Q_{\Lambda}({x})=0\}$, specifies the nearest-neighbor decoding region. The Voronoi cell is one example of the fundamental region of the lattice, which is defined as a measurable set $\mathcal{R}(\Lambda)\subset\mathbb{R}^{n}$ if $\cup_{\lambda\in\Lambda}(\mathcal{R}(\Lambda)+\lambda)=\mathbb{R}^{n}$ and if $(\mathcal{R}(\Lambda)+\lambda)\cap(\mathcal{R}(\Lambda)+\lambda')$ has measure 0 for any $\lambda\neq\lambda'$ in $\Lambda$. The modulo lattice operation can be defined with respect to (w.r.t.) a fundamental region $\mathcal{R}$ as $x \text{ mod}_{\mathcal{R}} \Lambda\triangleq {x}-Q_{\mathcal{R}}({x})$, where $Q_{\mathcal{R}}({x})$ represents a lattice quantizer according to the region $\mathcal{R}$. The volume of a fundamental region is equal to that of the Voronoi region $\mathcal{V}(\Lambda)$, which is given by $V(\Lambda)=|\text{det}({\mathbf{B}})|$. The volume-to-noise ratio (VNR) of an $n$-dimensional lattice $\Lambda$ is defined as $\gamma_{\Lambda}(\sigma)\triangleq V(\Lambda)^\frac{2}{n}/\sigma^2$. The NSM of $\Lambda$ is defined as $G(\Lambda) \triangleq \frac{1}{n V(\Lambda)}\cdot\frac{\int_{\mathcal{V}(\Lambda)}\|\mathbf{u}\|^2 d\mathbf{u}}{V^{2/n}(\Lambda)}$, where $\mathbf{u}$ is uniform in $\mathcal{V}(\Lambda)$.

\begin{deft}\label{deft:QZgood}
A sequence $\Lambda_n$ of lattices is called good for quantization or quantization-good under the mean square distortion measure if the NSM of $\Lambda_n$ satisfies
\begin{eqnarray}
\lim_{n \to \infty} G(\Lambda_n) = \frac{1}{2\pi e}.
\end{eqnarray}
\end{deft}

A sublattice $\Lambda' \subset \Lambda$ induces a partition (denoted by $\Lambda/\Lambda'$) of $\Lambda$ into equivalence groups modulo $\Lambda'$. The order of the partition, denoted by $|\Lambda/\Lambda'|$, is equal to the number of cosets. If $|\Lambda/\Lambda'|=2$, we call this a binary partition. Let $\Lambda(\Lambda_0)/\Lambda_{1}/\cdots/\Lambda_{r-1}/\Lambda' (\Lambda_{r})$ for $r > 1$ be an $n_p$-dimensional lattice partition chain. The following construction is known as ``Construction D" \cite[p.232]{yellowbook}. For each
partition $\Lambda_{\ell-1}/\Lambda_{\ell}$ ($1\leq \ell \leq r$) a code $C_{\ell}$ over $\Lambda_{\ell-1}/\Lambda_{\ell}$ selects a sequence of coset representatives $a_{\ell}$ in a set $A_{\ell}$ of representatives for the cosets of $\Lambda_{\ell}$. This construction requires a set of nested linear binary codes $C_\ell$ with block length $N$ and dimension of information bits $K_{\ell}$. When $\{C_1,...,C_r\}$ is a series of nested polar codes, we obtain a polar lattice \cite{yan2}. Note that the dimension of the constructed polar lattice is $n = n_p N$. Let $\psi$ be the natural embedding of $\mathbb{F}_{2}^{N}$ into $\mathbb{Z}^{N}$, where $\mathbb{F}_{2}$ is the binary field. Consider $\mathbf{g}_{1},\mathbf{g}_{2},\cdots,\mathbf{g}_{N}$ be a basis of $\mathbb{F}_{2}^{N}$ such that $\mathbf{g}_{1},\cdots\mathbf{g}_{k_{\ell}}$ span $C_{\ell}$, where $k_\ell$ is the dimension of $C_{\ell}$. When $n_p=1$, the binary lattice $L$ of Construction D consists of all vectors of the form
\begin{eqnarray}\label{eqn:latticeForm}
\sum_{\ell=1}^{r}2^{\ell-1}\sum_{j=1}^{k_{\ell}}u_{\ell}^{j}\psi(\mathbf{g}_{j})+2^{r}\mathbf{z},
\end{eqnarray}where $u_{\ell}^{j}\in\{0,1\}$ and $\mathbf{z}\in\mathbb{Z}^{N}$.

\subsection{Discrete Gaussian Distribution and Flatness Factor}
For $\sigma>0$ and $c\in\mathbb{R}^{n}$, recall that the standard Gaussian distribution of variance $\sigma^{2}$ centered at $c$ is defined by
\begin{eqnarray}
f_{\sigma,c}(x)\triangleq\frac{1}{(\sqrt{2\pi}\sigma)^{n}}e^{-\frac{\| x-c\|^{2}}{2\sigma^{2}}}, \:\:x\in\mathbb{R}^{n}.
\end{eqnarray}

Let $f_{\sigma,0}(x)=f_{\sigma}(x)$ for short. The differential entropy of $f_{\sigma}(x)$ is denoted by $h(\sigma^2)$. For an AWGN channel with noise variance $\sigma^2$ per dimension, the probability of error $P_e(\Lambda, \sigma^2)$ of a minimum-distance decoder for $\Lambda$ is
\begin{eqnarray}
P_e(\Lambda, \sigma^2)=1-\int_{\mathcal{V}(\Lambda)} f_{\sigma}(x) dx.
\end{eqnarray}

The $\Lambda$-periodic function is defined as
\begin{eqnarray}\label{eq:LamP}
f_{\sigma,\Lambda}(x)\triangleq\sum\limits_{\lambda\in\Lambda}f_{\sigma,\lambda}(x)=\frac{1}{(\sqrt{2\pi}\sigma)^{n}}\sum\limits_{\lambda\in\Lambda}e^{-\frac{\| x-\lambda\|^{2}}{2\sigma^{2}}}.
\end{eqnarray}

The discrete Gaussian distribution over $\Lambda$ centered at $c$ is defined as $D_{\Lambda, \sigma,c}(\lambda) \triangleq \frac{f_{\sigma,c}(\lambda)} {f_{\sigma,\Lambda}(c)}$, which is denoted by $D_{\Lambda, \sigma}(\lambda)$ when $c=0$. We note that $f_{\sigma,\Lambda}(x)$ is a probability density function (PDF) if $x$ is restricted to a fundamental region $\mathcal{R}(\Lambda)$. This distribution is actually the PDF of the $\Lambda$-aliased Gaussian noise, i.e., the Gaussian noise after the mod-$\Lambda$ operation \cite{forney6}. In this sense, the differential entropy of the $\Lambda$-aliased Gaussian noise is defined by
\begin{eqnarray}
h(\Lambda, \sigma^2) \triangleq -\int_{\mathcal{V}(\Lambda)} f_{\sigma,\Lambda}(x) \log f_{\sigma,\Lambda}(x) dx.
\end{eqnarray}
The flatness factor associated with $\Lambda$ is defined as follows to measure the difference between the distribution $f_{\sigma,\Lambda}(x)$ and the uniform distribution in $\mathcal{V}(\Lambda)$.
\begin{eqnarray}
\epsilon_{\Lambda}(\sigma)\triangleq\max\limits _{x\in\mathcal{R}(\Lambda)}\left|V(\Lambda)f_{\sigma,\Lambda}(x)-1\right|.\ \label{eq:flatnessFactor}
\end{eqnarray}

For two probability density functions $f_X$ and $g_X$, their total variation (TV) distance is defined by the following.
\begin{eqnarray}
\mathbb{V}(f_{X},g_X) \triangleq \frac{1}{2} \int_{x}|f_X(x)-g_X(x)| dx.
\end{eqnarray}The integral can be replaced with summation for discrete RVs.

\section{Polar Lattice Quantizer $\Lambda_Q$}\label{sec:QZ_Lattice}
\subsection{Test Channel with Discrete Gaussian Reconstruction}\label{sec:test}
\begin{figure}[ht]
    \centering
    \includegraphics[width=7cm]{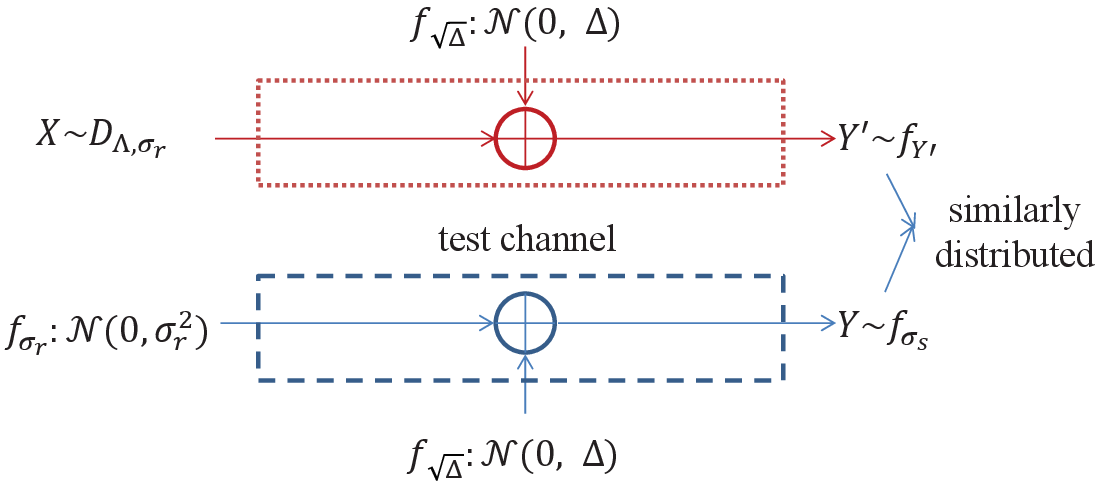}
    \caption{The genuine test channel (blue) for i.i.d. Gaussian sources and its approximate version (red) with discrete Gaussian reconstruction.}
    \label{fig:Test}
\end{figure}
In this section, we describe the construction of polar lattice quantizers and their quantization rule. To simplify the notations, we use the one-dimensional ($n_p=1$) binary partition chain ( $n_p=1$) $\eta\mathbb{Z}\cdots/\mathbb{Z}/\cdots/2^q\mathbb{Z}\cdots$ with a scaling factor $\eta$ in the following. Let $Y$ denote an i.i.d. Gaussian source with zero mean and variance $\sigma_s^2$, i.e., $Y\sim f_{\sigma_s}$. Consider the lossy compression of $Y$ with a target average squared-error distortion $\Delta \leq \sigma_s^2$. The test channel is an AWGN channel with noise variance $\Delta$, and the reconstruction is an i.i.d. Gaussian RV with variance $\sigma_r^2=\sigma_s^2-\Delta$, as depicted in the blue box in Fig. \ref{fig:Test}. However, a continuous reconstruction signal is impractical, and we turn to a lattice Gaussian distributed RV $X \sim D_{\Lambda, \sigma_r}$, as shown in the red box. After passing $X$ through the same AWGN channel, the channel output is a continuous RV $Y'\sim D_{\Lambda,\sigma_r}(\cdot) \circledast f_{\sqrt{\Delta}}(\cdot)$, where $\circledast$ denotes the standard convolution operation between two distributions. Our lattice quantizer can be designed based on $f_{Y'}$ instead of $f_{Y}$ thanks to the following result.

\begin{lem}[{\cite[Corollary 1]{LingBel13}}]\label{lem:TVYY'}
Let $\tilde{\sigma}^2_{\Delta}= \frac{\sigma_r^2\Delta}{\sigma_s^2}$. If $\epsilon_{\Lambda}(\tilde{\sigma}_{\Delta}) \leq \frac{1}{2}$, the TV distance between the density $f_{Y'}$ and the Gaussian density $f_Y$ satisfies $\mathbb{V}(f_{Y'},f_{Y}) \leq 2\epsilon_{\Lambda}(\tilde{\sigma}_{\Delta})$.
\end{lem}

Let $\eta\mathbb{Z}\cdots/\mathbb{Z}/\cdots/2^q\mathbb{Z}\cdots$ be the scaled one-dimensional binary partition chain. The quotient group is indexed by $X_\ell \in \{0, 1\}$ for each partition level. Then, a lattice Gaussian RV $X \sim D_{\Lambda,\sigma_r}$ can be uniquely expressed by the sequence $X_1, \dots, X_r, \dots$. The following proposition will be frequently needed since we require an exponentially vanishing flatness factor. The overall settings are illustrated in Fig. \ref{fig:1Dpartition}.
\begin{prop}[{\cite[Prop. 1]{LingQZ}}]\label{prop:morelevel}
Given a one-dimensional binary partition chain as in Fig. \ref{fig:1Dpartition}, for any $\tilde{\sigma}_\Delta$, scaling $\frac{1}{\eta^2}=O(N)$ guarantees an exponentially vanishing flatness factor $\epsilon_{\Lambda}(\tilde{\sigma}_\Delta)=O\left(e^{-N}\right)$. Moreover, using the first $r$ partition levels incurs a capacity loss $\frac{1}{2}\log \left(\frac{\sigma_s^2}{\Delta}\right)-I(X_{1:r};Y')= O\left(e^{-N}\right)$, if one chooses $2^q= \sqrt{N}$ and $r=\log\left(\frac{2^q}{\eta}\right)=O(\log N)$.
\end{prop}
\begin{IEEEproof}
See Appendix \ref{app:flatness} or \cite[Prop. 1]{LingQZ}.
\end{IEEEproof}

\begin{figure}[ht]
    \centering
    \begin{equation}
    \underbrace{\overbrace{\eta\mathbb{Z}(\Lambda)/\cdots/}^{\log(\frac{1}{\eta}) \text{ levels}}\mathbb{Z}\overbrace{/\cdots/2^{q}\mathbb{Z}(\Lambda')}^{q\text{ levels}}}_{r = \log(\frac{2^q}{\eta}) \text{ levels}}/\cdots
    \end{equation}
    \caption{The settings of the one-dimensional binary partition chain, where we choose $\frac{1}{\eta^2}= O(N)$, $2^q= \sqrt{N}$, and $r = q -\log(\eta) = O(\log N)$. }
    \label{fig:1Dpartition}
\end{figure}

Now we describe the construction of the polar lattice quantizer $\Lambda_Q$. An $N$ dimensional polar lattice quantizer works in a level-by-level manner, i.e., it compresses $Y'^{[N]}$ to $X_1^{[N]}$, then to $X_2^{[N]}$ based on $(X_1^{[N]},Y'^{[N]})$, and ends at the $r$-th level when $X_r^{[N]}$ is obtained based on $(X_{1:r-1}^{[N]},Y'^{[N]})$. The channel $X_r \to (Y', X_{1:r-1})$ is called the partition test channel at the level $r$ in \cite{LingQZ}, and its channel transition probability density function can be derived from $P_X$ and $f_{Y'|X}$. (cf. \cite[eq. (17)]{LingQZ}.) For the $\ell$-th level, let $U_{\ell}^{[N]}=X_{\ell}^{[N]}G_{N}$. We define the information set $\mathcal{I}_{\ell}$ and frozen set $\mathcal{F}_{\ell}$ as follows.
\begin{eqnarray}\label{eqn:BitAssign}
\mathcal{I}_{\ell}=\left\{i\in[N]:Z\left(U_{\ell}^{i}|U_{\ell}^{1:i-1},X_{1:\ell-1}^{[N]},Y'^{[N]}\right)< 1-2^{-N^{\beta}}\right\} \notag
\end{eqnarray} and $\mathcal{F}_{\ell}= \mathcal{I}^c_{\ell}$, where $Z\left(U_{\ell}^{i}|U_{\ell}^{1:i-1},X_{1:\ell-1}^{[N]},Y'^{[N]}\right)$ can be evaluated based on the $\ell$-th level $\Lambda_{\ell-1}/\Lambda_{\ell}$ partition channel with minimum mean square error (MMSE) re-scaled noise variance $\tilde{\sigma}_\Delta^2$ \cite[Lem. 10]{polarlatticeJ}. For a given realization $y^{[N]}$ of $Y'^{[N]}$, when the first $\ell-1$ levels complete, $x_{1:\ell-1}^{[N]}$ is recovered from $u_{1:\ell-1}^{[N]}$ using $G_N^{-1} = G_N$, the quantizer determines $u_{\ell}^{[N]}$ at the $\ell$-th level according to the following rule:
\begin{eqnarray}
& u_{\ell}^{i}=\begin{cases}
\begin{aligned} & 0\,\,\text{w. p.}\,\,\,P\left(0|u_{\ell}^{1:i-1},x_{1:\ell-1}^{[N]},y^{[N]}\right)\\
 & 1\,\,\text{w. p.}\,\,\,P\left(1|u_{\ell}^{1:i-1},x_{1:\ell-1}^{[N]},y^{[N]}\right)
\end{aligned}
\,\,\,\text{if}\,\,\,i\in\mathcal{I}_{\ell},\end{cases}\label{eqn:lossyencoder5}
\end{eqnarray}
and
\begin{eqnarray}
& u_{\ell}^{i}=\bar{u}_{\ell}^{i}\,\,\,\,\,\text{if}\,\,\,\,\,i\in\mathcal{F}_{\ell},
\label{eqn:lossyencoder6}
\end{eqnarray} where $\bar{u}_{\ell}^{i}$ is a pre-shared uniformly random bit.

\begin{figure}[ht]
    \centering
    \includegraphics[width=4cm]{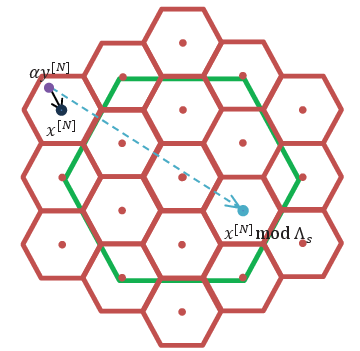}
    \caption{The lattice quantization processes with (dashed cyan line) and without (solid black line) shaping integrated, which correspond to the quantization rule in our work and that in \cite{LingQZ}, respectively.}
    \label{fig:QZnested}
\end{figure}

\begin{rem}
We note that the quantization rules \eqref{eqn:lossyencoder6} are different from that in \cite[eq. (24)]{LingQZ}, where a shaping set $\mathcal{S}_\ell$ is separately defined in $\mathcal{F}^c_\ell$. The reason for ignoring the shaping operation here is to better analyze the property of the underlying lattice quantizer. We note that the likelihood ratio for decoding is a function of $\alpha y^{[N]}$ due to the equivalence lemma \cite[Lem. 10]{polarlatticeJ}, where $\alpha = \sigma_r^2/\sigma_s^2$ is the MMSE re-scaling factor. It also means that the MAP decoding of $Y'^{[N]}$ w.r.t. a lattice Gaussian input $X^{[N]}$ is equivalent to the MMSE lattice decoding for $\alpha Y'^{[N]}$ \cite[Prop. 3]{LingBel13}. As can be seen from Fig. \ref{fig:QZnested}, for a scaled source realization $\alpha y^{[N]}$, our quantizer maps it to a lattice point (black dot) close to it. We note that the quantization lattice $\Lambda_Q$ (red in Fig. \ref{fig:QZnested}) may be shifted due to the random choices of $u^{\mathcal{F}_\ell}_\ell$. However, the lattice structure is not changed. When shaping is taken into consideration, the discrete Gaussian distribution of the reconstruction $X^{[N]}$ indeed forms a shaping lattice $\Lambda_s$ (green in Fig. \ref{fig:QZnested}), as has been proven in our recent work \cite{LingISIT2024}. Therefore, the polar lattice quantizer in \cite{LingQZ}, with discrete Gaussian shaping integrated, maps $\alpha y^{[N]}$ to $x^{[N]} \mod \Lambda_s$, as shown by the dashed cyan line in Fig. \ref{fig:QZnested}. It can be seen that the shaping operation induces slightly larger distortion, due to the rare probability of $\alpha y^{[N]}$ escaping $\mathcal{V}(\Lambda_s)$.
\end{rem}

\begin{ther}\label{thm:TVlvlr}
Let $Q_{U_{1:\ell}^{[N]},Y'{[N]}}$ denote the resulted joint distribution of $U_{1:\ell}^{[N]}$ and $Y'^{[N]}$ according to the encoding rules \eqref{eqn:lossyencoder5} and \eqref{eqn:lossyencoder6} for the first $\ell$ partition levels. Let $P_{U_{1:\ell}^{[N]},Y'^{[N]}}$ denote the joint distribution directly from $P_{X_{1:\ell} ,Y'}$, i.e., $U_{j}^{i}$ is generated according to the encoding rule \eqref{eqn:lossyencoder5} for all $i \in [N]$ and $j \leq \ell$. The TV distance between $P_{U_{1:\ell}^{[N]},Y'^{[N]}}$ and $Q_{U_{1:\ell}^{[N]},Y'^{[N]}}$ is upper-bounded as follows.
\begin{eqnarray}
\mathbb{V}\left(P_{U_{1:\ell}^{[N]},Y'^{[N]}},Q_{U_{1:\ell}^{[N]},Y'^{[N]}}\right) \leq \ell N\sqrt{\ln2\cdot2^{-N^\beta}}.
\end{eqnarray}
Particularly, when $r$ is chosen such that $P_e(\Lambda_r, \tilde{\sigma}_\Delta) \to 0$,
 \begin{eqnarray}
\mathbb{V}\left(P_{X^{[N]},Y'^{[N]}},Q_{X^{[N]},Y'^{[N]}}\right) \leq r N\sqrt{\ln2\cdot2^{-N^\beta}},
\end{eqnarray}where $Q_{X^{[N]},Y'^{[N]}}$ denotes the joint distribution between the (shifted) lattice point $X^{[N]}$ after quantization and the source $Y'^{[N]}$, and $P_{X^{[N]},Y'^{[N]}}$ denotes that between the input and output of $N$ copies of the approximate test channel in Fig. \ref{fig:Test}.
\end{ther}
\begin{IEEEproof}
See Appendix \ref{app:TVlvlr} or \cite[Thm. 3, eq. (26)]{LingQZ}.
\end{IEEEproof}

\subsection{Performance of The Polar Lattice Quantizer $\Lambda_Q$}\label{sec:QZPerf}
When $\Lambda_Q$ is applied to the genuine Gaussian source $Y \sim f_{\sigma_s}$, we have the following lemma as a consequence of Lemma \ref{lem:TVYY'}. The proof can be obtained by using the triangle inequality and the fact that $\mathbb{V}(f_{Y^{[N]}}, f_{Y'^{[N]}}) \leq 2N\epsilon_{\Lambda}(\tilde{\sigma}_{\Delta})$.
\begin{lem}\label{lem:TVTogether}
Let $Q_{X^{[N]},Y^{[N]}}$ denote the joint distribution between the source vector $Y^{[N]}$ and the reconstruction $X^{[N]}$ when $\Lambda_Q$ (with shift) is applied to the i.i.d. Gaussian source. The TV distance between $Q_{X^{[N]},Y^{[N]}}$ and $P_{X^{[N]},Y'^{[N]}}$ satisfies
\begin{eqnarray}
\begin{aligned}
\mathbb{V}&\left(Q_{X^{[N]},Y^{[N]}},P_{X^{[N]},Y'^{[N]}}\right) \\
&\leq r N\sqrt{\ln2\cdot2^{-N^\beta}} + 2N\epsilon_{\Lambda}(\tilde{\sigma}_{\Delta}) \\
&\leq 2\cdot2^{-N^{\beta'}},
\end{aligned}
\end{eqnarray}for some constant $0<\beta'<\beta<\frac{1}{2}$ and sufficiently large $N$.
\end{lem}

We are now interested in the average quadratic distortion achieved by $\Lambda_Q$. It is easier to evaluate the average quadratic distortion $\mathsf{E}_{P_{X^{[N]},Y'^{[N]}}}\left[\big\|\alpha Y'^{[N]}-X^{[N]}\big\|^2\right]$ (shorted as $\mathsf{E}_{P}\left[\big\|\alpha Y'^{[N]}-X^{[N]}\big\|^2\right]$) under the joint distribution $P_{X^{[N]},Y'^{[N]}}$ instead. In the ideal case that both $X$ and $Y$ are Gaussian RVs, $\alpha Y - X$ is a Gaussian RV with variance $\tilde{\sigma}_\Delta^2$. The proofs of Lemma \ref{lem:EPXY'} and Theorem \ref{thm:EDistor} are given in Appendix \ref{app:EPXY'} and \ref{app:EDistor}, respectively.
\begin{lem}\label{lem:EPXY'}
Let $X^{[N]}$ and $Y'^{[N]}$ be drawn from $N$ i.i.d. copies of the approximate test channel given in Fig. \ref{fig:Test}. The average distortion between $\alpha Y'^{[N]}$ and $X^{[N]}$ per dimension satisfies
\begin{eqnarray}
\left|\frac{1}{N}\mathsf{E}_{P}\left[\big\|\alpha Y'^{[N]}-X^{[N]}\big\|^2\right]-\tilde{\sigma}^2_\Delta \right|\leq \frac{2\pi\epsilon_1}{1-\epsilon_1}\frac{\Delta}{\sigma_s^2}\cdot\tilde{\sigma}^2_\Delta,
\end{eqnarray}where $\epsilon_1 \triangleq \epsilon_{\Lambda}\left(\sigma_r/\sqrt{\frac{\pi}{\pi-1/e}}\right)$.
\end{lem}

\begin{rem}
Since $\tilde{\sigma}^2_\Delta = \frac{\Delta\sigma_r^2}{\sigma_s^2} \leq \sigma_r^2$ and $\sqrt{\frac{\pi}{\pi-1/e}} \approx 1.06$, the condition $\epsilon_1 \to 0$ can be easily handled when $\epsilon_{\Lambda}(\tilde{\sigma}_\Delta) \to 0$.
\end{rem}

\begin{ther}\label{thm:EDistor}
For the i.i.d. Gaussian source vector $Y^{[N]}$ and its reconstruction $X^{[N]}$ after the quantization of $\Lambda_Q$ (with shift), the average quadratic distortion per dimension can be upper bounded as follows.
\begin{eqnarray}
\frac{1}{N} \mathsf{E}_{Q}\left[\big\|\alpha Y^{[N]}-X^{[N]}\big\|^2\right] \leq \tilde{\sigma}^2_\Delta + 2N\cdot2^{-N^{\beta'}}.
\end{eqnarray}
\end{ther}

\begin{rem}\label{rem:compare}
As shown in Fig. \ref{fig:QZnested}, removing the shaping operation in the quantization process gives us more convenience on the bounding $\left\|\alpha Y^{[N]}-X^{[N]}\right\|^2$. Since $2^q\mathbb{Z}^N \subseteq \Lambda_Q$, $\left\|\alpha Y^{[N]}-X^{[N]}\right\|^2$ is bounded by $N\cdot2^{q-1}$. When shaping is integrated, the upper bound of the average quadratic distortion includes two extra terms, which are caused by the two cases when $X^{[N]}$ and $\alpha Y^{[N]}$ escape the shaping region of $\Lambda_s$ (green hexagon in Fig. \ref{fig:QZnested}), as summarized in \cite[eq. (65)]{LingQZ}.
\end{rem}

We then prove that the variance of $\frac{1}{N}\big\|\alpha Y^{[N]}-X^{[N]}\big\|^2$ converges to 0 as $N$ grows. Before that, we show that the fourth moment of the discrete Gaussian RV is close to that of the continuous Gaussian RV when the flatness factor is small. The following lemma is a direct generalization of \cite[Lem. 6]{cong2} and \cite[Lem. 5]{LingBel13}. Although the result may be extended to $n$-dimensional lattices, we focus on the simplest case where $n=1$ for brevity.
\begin{lem}\label{lem:4thGap}
Let $X$ be sampled from the Gaussian distribution $D_{\Lambda, \sigma_r, c}$, where $\Lambda$ is a one-dimensional lattice and $c$ is a scalar. If $\epsilon_2 \triangleq \epsilon_{\Lambda}\left(\sigma_r/\sqrt{\frac{\pi}{\pi-2/e}}\right) \leq 1$, then
\begin{eqnarray}
\Big|\mathsf{E}\left[\|X-c\|^4\right] - 3\sigma_r^4 \Big| \leq 4(\pi+3)\frac{\epsilon_2}{1-\epsilon_2}\sigma_r^4.
\end{eqnarray}
\end{lem}
\begin{IEEEproof}
See Appendix \ref{app:4thGap}.
\end{IEEEproof}

Roughly speaking, for the approximate test channel, $\alpha Y' - X$ is close to an i.i.d. Gaussian RV, then the variance of $\frac{1}{N}\big\|\alpha Y'^{[N]}-X^{[N]}\big\|^2$ under distribution $P_{X^{[N]}, Y'^{[N]}}$ converges to 0 as $N$ increases. This result also holds for $Q_{X^{[N]}, Y^{[N]}}$, by Lemma \ref{lem:TVTogether}. The proof of Theorem \ref{thm:VarDistor} is given in Appendix \ref{app:VarDistor}.
\begin{ther}\label{thm:VarDistor}
For the i.i.d. Gaussian source vector $Y^{[N]}$ and its reconstruction $X^{[N]}$ after the quantization of $\Lambda_Q$ (shifted), the variance of the quadratic distortion per dimension can be upper-bounded as follows.
\begin{eqnarray}
\mathsf{Var}_Q\left[\frac{1}{N}\big\|\alpha Y^{[N]}-X^{[N]}\big\|^2\right] \leq N^2 \cdot 2^{-N^{\beta'}}.
\end{eqnarray}
\end{ther}

\begin{figure}[ht]
    \centering
    \includegraphics[width=5cm]{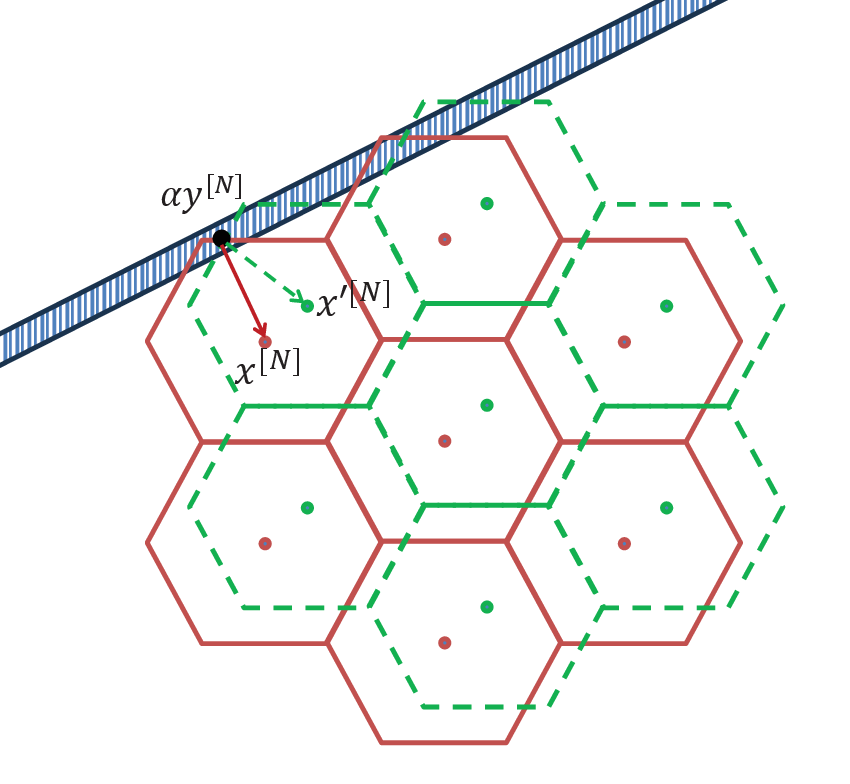}
    \caption{The quantization of source $\alpha Y^{[N]}$ with shifted fine lattices. When the shape of $\mathcal{V}(\Lambda_Q)$ is not sharp, as some of its corners being merged in the spherical shell, $\big\|\alpha y^{[N]}-x^{[N]}\big\|^2 \approx \big\|\alpha y^{[N]}-x'^{[N]}\big\|^2$.}
    \label{fig:QZshift}
\end{figure}

\begin{rem}
We can show that the above variance approaches 0 as $N \to \infty$, and therefore $\frac{1}{N}\big\|\alpha Y^{[N]}-X^{[N]}\big\|^2$ converges to $\frac{1}{N} \mathsf{E}_{Q}\left[\big\|\alpha Y^{[N]}-X^{[N]}\big\|^2\right]$ in probability. We note that since the frozen sets (see \eqref{eqn:lossyencoder6}.) of the polar codes are filled with random bits (rather than all-zeros), we actually utilize a coset $\Lambda_Q+{c}'$ of the polar lattice for the quantization of $\alpha y^{[N]}$, where the shift ${c}'$ accounts for the effects of random frozen bits. One may think that the average quadratic distortion $\frac{1}{N}\big\|\alpha Y^{[N]}-X^{[N]}\big\|^2$ will be dramatically changed if the underlying lattice is randomly shifted. However, Theorem \ref{thm:VarDistor} tells us that the fluctuation in $\frac{1}{N}\big\|\alpha Y^{[N]}-X^{[N]}\big\|^2$ is vanishing as $N$ increases. This can be better understood by fixing the quantization lattice and increasing the Gaussian source variance $\sigma_s^2$, i.e., by fixing $\tilde{\sigma}^2_\Delta$ and setting $\Delta = \frac{\sigma_s^2-\sqrt{\sigma_s^4-4\sigma_s^2\tilde{\sigma}^2_\Delta}}{2}$ for any given large $\sigma_s^2$. It is well known that the high-dimensional Gaussian distribution turns to a spherical shell around the origin \cite{BK:Zamir}, and most of its volume is on the surface. When $\sigma_s^2$ is relatively large ($\alpha \to 1$) compared with $\tilde{\sigma}_\Delta^2$, the curve of the spherical shell corresponding to the source vector is approximately straight w.r.t. the quantization lattice around the boundary, as depicted in Fig. \ref{fig:QZshift}. Theorem \ref{thm:VarDistor} implies that the shape of the shaping region of the lattice $\Lambda_Q$ is not sharp and $\big\|\alpha Y^{[N]}-X^{[N]}\big\|^2$ is insensitive to the shift of $\Lambda_Q$.
\end{rem}

Combining this with Theorem \ref{thm:EDistor}, we see that the quantization noise per dimension is close to $\tilde{\sigma}^2_\Delta$ with high probability, which yields an upper bound on the second moment of $\Lambda_Q$.
\begin{lem}\label{lem:SecMBound}
Let the second moment of $\Lambda_Q$ be denoted by $\sigma^2(\Lambda_Q) \triangleq \frac{\int_{\mathcal{V}(\Lambda_Q)}\|\mathbf{u}\|^2 d\mathbf{u}}{NV(\Lambda_Q)}$, where $\mathbf{u}$ is uniform in $\mathcal{V}(\Lambda_Q)$. Then, for some constant $0<\beta'' < \beta' < \frac{1}{2}$ and sufficiently large $N$,
\begin{eqnarray}
\sigma^2(\Lambda_Q) \leq \frac{N}{N+2} \left(\tilde{\sigma}^2_\Delta + 2^{-N^{\beta''}}\right).
\end{eqnarray}
\end{lem}
\begin{IEEEproof}
See Appendix \ref{app:SecMBound}.
\end{IEEEproof}

\section{The Quantization-goodness of Polar Lattices}\label{sec:EQshap}
In this section, we first investigate the volume $V(\Lambda_Q)$ of the quantization lattice. This idea is similar to that used for the AWGN-good polar lattices in \cite{polarlatticeJ}, where we proved that the VNR of a polar lattice $\Lambda_c$ designed for reliable transmission over the AWGN channel with noise variance $\sigma^2$ and without power constraint is close to $2\pi e$. For the quantization polar lattice $\Lambda_Q$, we remind the readers that its construction is based on the unlimited-input AWGN channel with noise variance $\tilde{\sigma}^2_\Delta$. Using the capacity-achieving property of polar codes, one may also expect that $\gamma_{\Lambda_Q}(\tilde{\sigma}_\Delta) \to 2\pi e$.
\begin{lem}\label{lem:VNRbound}
For the polar lattice $\Lambda_Q$ constructed from $r$ component polar codes according to $\mathcal{I}_\ell$ for each level and the binary partition chain in Fig. \ref{fig:1Dpartition}, the VNR of $\Lambda_Q$ w.r.t. variance $\tilde{\sigma}^2_\Delta$ can be lower bounded as follows.
\begin{eqnarray}
\log\left(\frac{\gamma_{\Lambda_Q}(\tilde{\sigma}_\Delta)}{2\pi e}\right) = 2 (\epsilon_a - \epsilon_b - \epsilon_c) \geq -2(\epsilon_b + \epsilon_c),
\end{eqnarray}where $\epsilon_a = C(\Lambda, \tilde{\sigma}^2_\Delta)$, $\epsilon_b \triangleq h(\tilde{\sigma}^2_\Delta)-h(\Lambda', \tilde{\sigma}^2_\Delta)$ and $\epsilon_c \triangleq \sum_{\ell=1}^r R_\ell - C(\Lambda_{\ell-1}/\Lambda_{\ell}, \tilde{\sigma}^2_\Delta)$ for coding rate $R_\ell = \frac{|\mathcal{I}_\ell|}{N}$.
\end{lem}

\begin{rem}
$C(\Lambda,\tilde{\sigma}^2_\Delta)$ and $C(\Lambda_{\ell-1}/\Lambda_{\ell}, \tilde{\sigma}^2_\Delta)$ are the capacities of the mod-$\Lambda$ channel and the $\Lambda_{\ell-1}/\Lambda_{\ell}$ channel, respectively, defined in \cite{forney6}. Roughly speaking, by appropriately scaling the partition chain, one can make $\eta$ small and $\Lambda$ is sufficiently fine such that $\epsilon_a \approx 0$. In fact, one can prove that $C(\Lambda, \tilde{\sigma}^2_\Delta) \leq \log(e)\cdot \epsilon_{\Lambda}(\tilde{\sigma}_\Delta)$, using the definition of the flatness factor. The proof is skipped here since $\epsilon_a \geq 0$ is enough for the lower bound. By increasing $r$, $\Lambda'$ can be made coarse enough compared with $\tilde{\sigma}^2_\Delta$ such that the $\text{mod-} \Lambda'$ operation has a little influence on the Gaussian noise, making $\epsilon_b \approx 0$. Finally, by the property of channel polarization \cite{arikan2009channel}, $R_\ell$ approaches $C(\Lambda_{\ell-1}/\Lambda_{\ell}, \tilde{\sigma}^2_\Delta)$ as $N$ grows, which yields $\epsilon_c \approx 0$. As a result, we see that $V(\Lambda_Q)^{\frac{2}{N}} \approx 2\pi e \tilde{\sigma}^2_\Delta$ from the above equality.
\end{rem}

We are now ready to address our main theorem of this work, which is a direct result of Lemma \ref{lem:SecMBound} and Lemma \ref{lem:VNRbound}. See Appendix \ref{app:QZgood} for its proof.
\begin{ther}\label{thm:QZgood}
For the polar lattice $\Lambda_Q$ constructed from $r$ component polar codes according to $\mathcal{I}_\ell$ for each level and the binary partition chain in Fig. \ref{fig:1Dpartition}, where the top lattice $\Lambda = \eta \mathbb{Z}$ and the bottom lattice $\Lambda'=\eta 2^r \mathbb{Z}$, the NSM of $\Lambda_Q$ can be upper bounded as
\begin{eqnarray}
G(\Lambda_Q) \leq \frac{N}{N+2} 2^{2(\epsilon_b +\epsilon_c)} \frac{1}{2\pi e} \left(1+\frac{2^{-N^{\beta''}}}{\tilde{\sigma}^2_\Delta}\right).
\end{eqnarray}
$\Lambda_Q$ is quantization good in the sense that $\underset{N\to \infty}{\lim} G(\Lambda_Q) = \frac{1}{2\pi e}$. Moreover, the rate of quantization goodness can be characterized as
\[
\lim_{N\to \infty}\log(G(\Lambda_Q)\cdot 2\pi e) = O\left(\frac{\log N}{N^\frac{1}{\mu}}\right),
\]where $\mu$ is the scaling factor of polar codes \cite{HassaniScal14}.
\end{ther}

\begin{rem}
It should be noted that there is a trade-off between $\beta''$ and $\mu$ in the finite scaling of binary polar codes. One can only have $0<\beta''<\beta<\frac{1}{2}$, when the overhead $\epsilon_c>0$ is fixed. Similarly, the lower bound $\mu>4.63$ is valid when the decoding error probability of polar codes is fixed. To guarantee a valid construction of polar lattices according to the threshold $1-2^{-N^{\beta}}$ while maintaining a positive scaling factor $\mu$, we turn to the joint scaling issue of error probability and gap to capacity, which has been studied in \cite[Sect. IV]{MondelliScal16}. In \cite[Theorem 3]{MondelliScal16}, the block error probability of polar codes under successive cancellation decoding and the block length are characterized as the following scaling rules.
\begin{eqnarray}
P_e \leq N\cdot 2^{-N^{\gamma\cdot h_2^{(-1)}\left(\frac{\gamma(\mu+1)-1}{\gamma\mu}\right)}}
\end{eqnarray} 
and 
\begin{eqnarray}
N \leq \frac{\beta_u}{|C(W)-R|^{\mu/(1-\gamma)}},
\end{eqnarray}
where $\gamma \in (1/(1+\mu),1)$, $\beta_u$ is a universal constant that does not depend of the test channel $W$ or on $\gamma$, and $h_2^{(-1)}(\cdot)$ is the inverse of the binary entropy function. In other words, $\beta''$ is upper-bounded by $\beta''<\beta< \gamma\cdot h_2^{(-1)}\left(\frac{\gamma(\mu+1)-1}{\gamma\mu}\right)$ and $\mu$ gets worsen by a factor $\frac{1}{1-\gamma}$. To obtain a good bound on the overhead of rate for a positive $\beta''$, one may choose $\gamma \to 1/(1+\mu)$, and the scaling of $N$ becomes $O(1/|C(W)-R|^{\mu+1})$. It is believed that the additional penalty $1$ on $\mu$ is only an artifact of the proof technique, and the scaling factor is still $\mu$. See \cite[Remark 4]{MondelliScal16} for more details.
\end{rem}

\begin{rem}
The scaling factor $\mu$ is characterized as 5.702 for both channel and source coding in \cite{GoldinScal14}. There are better choices of $\mu$ for polar codes when the underlying channel is binary erasure channel\cite{HassaniScal14}. A more recent result shows that $\mu$ can be improved to 4.714 for any BMSC \cite{MondelliScal16}. To the best of our knowledge, the latest record of $\mu$ is 4.63, as presented in \cite{HPWangScaling2023}. Let $G_N^*$ denote the NSM of an sphere of $N$ dimension, then $G_N^* \to \frac{1}{2 \pi e}$ with a rate of $\log(G_N^* \cdot 2\pi e ) = o \left(\frac{\log N}{N}\right)$ \cite{BK:Zamir}. For polar lattices, the best known rate is $O\left(\log N/{N^\frac{1}{4.63}}\right)$.
\end{rem}

\section{Frozen bits do not matter}\label{sec:Fzero}
In this section, our aim is to fix the frozen bits $\bar{u}_\ell^i$ for $i \in \mathcal{F}_\ell$ at each level (see \eqref{eqn:lossyencoder6}.) and consequently utilize the offset-fixed quantization lattice for better implementation. It is worth noting that this procedure does not alter the NSM of $\Lambda_Q$, and those readers solely interested in the quantization goodness of polar lattices can skip this part safely. Our strategy relies on treating the expectation $\mathsf{E}_{Q}\left[\big\|\alpha Y^{[N]}-X^{[N]}\big\|^2\right]$ as an average of performance over all uniformly random $u_\ell^{\mathcal{F}_\ell}$ and the random mappings in \eqref{eqn:lossyencoder5} for $1\leq \ell \leq r$. By further bounding the conditional variance of $\mathsf{E}_{Q}\left[\big\|\alpha Y^{[N]}-X^{[N]}\big\|^2\right]$ w.r.t. a given choice of $u_\ell^{\mathcal{F}_\ell}$, we demonstrate that our lattice quantizer with a fixed offset yields nearly identical performance to that based on a pre-shared random offset as described in Sect. \ref{sec:QZ_Lattice}.

Let $U^{\mathcal{F}_{\Lambda_Q}}=\{U_1^{\mathcal{F}_1},..., U_\ell^{\mathcal{F}_\ell},..., U_r^{\mathcal{F}_r}\}$ denote the collection of all the frozen bits over the $r$ levels. For a specific choice $u^{\mathcal{F}_{\Lambda_Q}}=\{u_1^{\mathcal{F}_1},..., u_\ell^{\mathcal{F}_\ell},..., u_r^{\mathcal{F}_r}\}$, the resulted in offset over $\Lambda_Q$ can be expressed as
\begin{eqnarray}
\eta\sum_{\ell=1}^{r}2^{\ell-1}\sum_{j=k_\ell+1}^Nu_{\ell}^{j}\psi(\mathbf{g}_{j}),\label{eqn:shift}
\end{eqnarray}
where $u_{\ell}^{j}\in\{0,1\}$ and $\mathbf{g}_{k_\ell+1},\cdots,\mathbf{g}_{N}$ are the bases that left in the row space of $G_N$ after $\mathbf{g}_{1},\cdots,\mathbf{g}_{k_\ell}$ are picked for $C_{\ell}$ as in \eqref{eqn:latticeForm}. $U^{\mathcal{F}_{\Lambda_Q}}$ can be flattened as a sequence consisting of $|\mathcal{F}_{\Lambda_Q}|=rN-\sum^r_{\ell=1}k_\ell = N(r-R_{\mathcal{C}})$ i.i.d. uniformly random bits. Recall that in Theorem \ref{thm:EDistor}, $\mathsf{E}_{Q}\left[\frac{1}{N}\big\|\alpha Y^{[N]}-X^{[N]}\big\|^2\right]$ represents the average performance of $\Lambda_Q$ on all random choices of $U^{\mathcal{F}_{\Lambda_Q}}$ and random mappings \eqref{eqn:lossyencoder5} over all $\mathcal{I}_\ell$. By the linearity of expectation, it can be written as $\mathsf{E}_{U^{\mathcal{F}_{\Lambda_Q}}}\left[\mathsf{E}_{Q|U^{\mathcal{F}_{\Lambda_Q}}}\left[\frac{1}{N}\big\|\alpha Y^{[N]}-X^{[N]}\big\|^2\right]\right]$, where $\mathsf{E}_{Q|U^{\mathcal{F}_{\Lambda_Q}}}\left[\frac{1}{N}\big\|\alpha Y^{[N]}-X^{[N]}\big\|^2\right]$ stands for the average distortion caused by using a specific offset corresponding to $U^{\mathcal{F}_{\Lambda_Q}}$. The following theorem states that $\mathsf{E}_{Q|U^{\mathcal{F}_{\Lambda_Q}}}\left[\frac{1}{N}\big\|\alpha Y^{[N]}-X^{[N]}\big\|^2\right]$ approaches $\mathsf{E}_{Q}\left[\frac{1}{N}\big\|\alpha Y^{[N]}-X^{[N]}\big\|^2\right]$ in probability as $N$ increases, and the result of Theorem \ref{thm:EDistor} can be applied to $\mathsf{E}_{Q|U^{\mathcal{F}_{\Lambda_Q}}}\left[\frac{1}{N}\big\|\alpha Y^{[N]}-X^{[N]}\big\|^2\right]$ with slight modification.

\begin{ther}\label{thm:fixF}
Let $U^{\mathcal{F}_{\Lambda_Q}}$ be the combination of all random frozen bits that determine the offset of $\Lambda_Q$ as in \eqref{eqn:shift}. For any specific choice $u^{\mathcal{F}_{\Lambda_Q}}$ of $U^{\mathcal{F}_{\Lambda_Q}}$, the performance of the offset-fixed lattice quantizer is good in the sense that
\begin{eqnarray}
\mathsf{Prob.}\bigg(\mathsf{E}_{Q|u^{\mathcal{F}_{\Lambda_Q}}}[*] \leq \tilde{\sigma}^2_\Delta &+ 2N2^{-N^{\beta'}}+N^{2/3}2^{-\frac{1}{3}N^{\beta'}}\bigg) \notag \\ 
&\geq 1-N^{2/3}2^{-\frac{1}{3}N^{\beta'}}, \notag
\end{eqnarray}where $*$ is short for $\frac{1}{N}\|\alpha Y^{[N]}-X^{[N]}\big\|^2$.
\end{ther}
\begin{IEEEproof}
See Appendix \ref{app:fixF}.
\end{IEEEproof}

\begin{rem}\label{rem:fixLattice}
The above theorem says that the average quantization noise per dimension is slightly larger than $\tilde{\sigma}^2_\Delta$ with high probability. In fact, Lemma \ref{lem:SecMBound} can be rewritten for the offset-fixed lattice quantizer and its second moment is upper-bounded in a similar way. To see this, we use the abbreviation $*$ for $\frac{1}{N}\left\|\alpha Y^{[N]} -X^{[N]}\right\|^2$ again. Let $\mathcal{E}_a$, $\mathcal{E}_b$ and $\mathcal{E}_c$ denote the events that $\mathsf{E}_{Q|U^{\mathcal{F}_{\Lambda_Q}}}[*] \geq \mathsf{E}_Q [*] + \delta_a$, $\mathsf{Var}_{Q|U^{\mathcal{F}_{\Lambda_Q}}}[*] \geq \delta_b$ and $\frac{1}{N}\left\|\alpha Y^{[N]} -X^{[N]}\right\|^2 \geq \mathsf{E}_{Q|U^{\mathcal{F}_{\Lambda_Q}}}[*]+\delta_c$, respectively. By Markov's inequality and Chebyshev's inequality, $\mathsf{Prob.}(\mathcal{E}_a \cup \mathcal{E}_b \cup \mathcal{E}_c) \leq \frac{\mathsf{Var}_{U^{\mathcal{F}_{\Lambda_Q}}}\left[\mathsf{E}_{Q|U^{\mathcal{F}_{\Lambda_Q}}}[*]\right]}{\delta_a^2}+\frac{\mathsf{E}_{U^{\mathcal{F}_{\Lambda_Q}}}\left[\mathsf{Var}_{Q|U^{\mathcal{F}_{\Lambda_Q}}}[*]\right]}{\delta_b}+\frac{\mathsf{Var}_{Q|U^{\mathcal{F}_{\Lambda_Q}}}[*]}{\delta_c^2}$. Then, we have \eqref{eqn:rmklong1}, which immediately yields \eqref{eqn:rmklong2}, where we choose $\delta_a^2=\delta_b$ and employ \eqref{eqn:TotalVar} again. Finally, letting $\delta_a=\sqrt[3]{\mathsf{Var}_Q[*]}$ and $\delta_c=\sqrt[6]{\mathsf{Var}_Q[*]}$ will result in a similar result as that given in Lemma \ref{lem:SecMBound}. We skip the rest of the proof since Lemma \ref{lem:SecMBound} is sufficient for the quantization goodness of $\Lambda_Q$, as we have mentioned.
\end{rem}

\begin{figure*}[th]
\begin{eqnarray}\label{eqn:rmklong1}
\mathsf{Prob.}(\bar{\mathcal{E}}_a \cap \bar{\mathcal{E}}_b \cap \bar{\mathcal{E}}_c) \geq 1- \left[\frac{\mathsf{Var}_{U^{\mathcal{F}_{\Lambda_Q}}}\left[\mathsf{E}_{Q|U^{\mathcal{F}_{\Lambda_Q}}}[*]\right]}{\delta_a^2}+\frac{\mathsf{E}_{U^{\mathcal{F}_{\Lambda_Q}}}\left[\mathsf{Var}_{Q|U^{\mathcal{F}_{\Lambda_Q}}}[*]\right]}{\delta_b}+\frac{\mathsf{Var}_{Q|U^{\mathcal{F}_{\Lambda_Q}}}[*]}{\delta_c^2}\right],
\end{eqnarray}

\begin{eqnarray}\label{eqn:rmklong2}
\begin{aligned}
\mathsf{Prob.}\left(\frac{1}{N}\left\|\alpha Y^{[N]} -X^{[N]}\right\|^2 \leq \mathsf{E}_Q[*] +\delta_a +\delta_c\right) &\geq 1- \left[\frac{\mathsf{Var}_{U^{\mathcal{F}_{\Lambda_Q}}}\left[\mathsf{E}_{Q|U^{\mathcal{F}_{\Lambda_Q}}}[*]\right]}{\delta_a^2}+\frac{\mathsf{E}_{U^{\mathcal{F}_{\Lambda_Q}}}\left[\mathsf{Var}_{Q|U^{\mathcal{F}_{\Lambda_Q}}}[*]\right]}{\delta_b}+\frac{\delta_b}{\delta_c^2}\right]\\
&= 1- \frac{\mathsf{Var}_Q[*]}{\delta_a^2}-\frac{\delta_b}{\delta_c^2},
\end{aligned}
\end{eqnarray}
\hrulefill
\end{figure*}

\section{Conclusion}
In this work, we prove that the polar lattices constructed for lossy compression are indeed quantization-good. We also demonstrate that the rate of quantization goodness of polar lattices is determined by the scaling factor of polar codes. For a more convenient implementation, we fix the offset of the quantization lattice, i.e. fix $\bar{u}^i_\ell $ in \eqref{eqn:lossyencoder6}, which means that dither is not needed for our quantization scheme.





%
\bibliographystyle{IEEEtran}
\bibliography{Myreff}

\newpage
\appendix
\section{Appendices}
\subsection{Proof of Proposition \ref{prop:morelevel}}\label{app:flatness}
\begin{IEEEproof}
For the one dimensional lattice partition chain, recall that the top lattice $\Lambda=\eta \mathbb{Z}$ for some scaling $\eta$. Let $\Lambda^*=\frac{1}{\eta}\mathbb{Z}$ be the dual lattice of $\Lambda$. By \cite[Corollary 1]{cong2}, using the alternative definition of theta series $\Theta_{\Lambda}(\tau)=\sum_{\lambda\in\Lambda}e^{-\pi\tau\|\lambda\|^2}$, we have
\allowdisplaybreaks{\begin{eqnarray}
\epsilon_{\Lambda}(\tilde{\sigma}_\Delta)&=&\Theta_{\Lambda^*}\left(2\pi \tilde{\sigma}_\Delta^2\right)-1 \\
&=&\sum_{\lambda \in \Lambda^*}\exp\left(-2\pi^2\tilde{\sigma}_\Delta^2\|\lambda\|^2 \right)-1\\
&=&2\sum_{\lambda \in \frac{1}{\eta}\mathbb{Z}_+} \exp\left(-2\pi^2\tilde{\sigma}_\Delta^2\|\lambda\|^2 \right) \\
&\leq&\frac{2\exp\left(-2 \pi^2\tilde{\sigma}_\Delta^2\frac{1}{\eta^2}\right)}{1-\exp\left(-2 \pi^2\tilde{\sigma}_\Delta^2\frac{3}{\eta^2}\right)}\\
&\leq&4\exp\left(-2 \pi^2\tilde{\sigma}_\Delta^2\frac{1}{\eta^2}\right),
\end{eqnarray}
where $\mathbb{Z}_{+}$ denotes positive integers and the last inequality satisfies for sufficiently small $\eta$.} Let $\frac{1}{\eta^2}=O(N)$ so that $\epsilon_{\Lambda}(\tilde{\sigma})=O\left(e^{-N}\right)$. According to \cite[Lem. 5]{polarlatticeJ}\footnote{Although this lemma proves that when $2^q = O(\log N)$, the capacity loss is $O(\frac{1}{N})$, it can be easily modified by fixing $2^q = \sqrt{N}$ so that the capacity loss is $O\left(e^{-N}\right)$.}, the partition chain with bottom lattice $\Lambda'=2^{q}\mathbb{Z}$ and $2^q=\sqrt{N}$ can guarantee a capacity loss $\sum_{\ell>r_1}I(Y';X_{\ell}|X_{1:\ell-1})= O\left(e^{-N}\right)$. Finally, the number of levels for partition $\eta\mathbb{Z}/\cdots/\mathbb{Z}/\cdots/2^{q}\mathbb{Z}$ satisfies $r=\log(\frac{2^{q}}{\eta})=O(\log N)$. Combining this with \cite[Thm. 2]{LingBel13}, it can be found that the rate of convergence to the rate-distortion bound is $O(e^{-N})$ by using the first $r$ partition levels. The overall settings of the partition chain is illustrated in Fig. \ref{fig:1Dpartition}.
\end{IEEEproof}

\subsection{Proof of Theorem \ref{thm:TVlvlr}}\label{app:TVlvlr}
\begin{IEEEproof}
We start with the first level by proving that
\begin{eqnarray}
\mathbb{V}\left(P_{U_{1}^{1:N},Y'^{1:N}},Q_{U_{1}^{1:N},Y'^{1:N}}\right) \leq N\sqrt{\ln2\cdot2^{-N^\beta}}.
\end{eqnarray}

Using the telescoping expansion
\begin{equation}\label{eq:telescp}
B^{[N]}-A^{[N]}=\sum_{i=1}^N(B^i-A^i)A^{1:i-1}B^{i+1:N},
\end{equation}$\mathbb{V}\left(P_{U_1^{[N]},Y'^{[N]}},Q_{U_1^{[N]},Y'^{[N]}}\right)$ can be decomposed as \eqref{eqn:longTV}, where $\mathbb{D}_1(\cdot||\cdot)$ is the Kullback-Leibler divergence, and the equalities and the inequalities follow from
\begin{itemize}
\item[] $(a)$ $Q\left(u_1^i|u_1^{1:i-1},y^{[N]}\right)=P\left(u_1^i|u_1^{1:i-1},y^{[N]}\right)$ for $i \in \mathcal{I}_1$.
\item[] $(b)$ Pinsker's inequality.
\item[] $(c)$ Jensen's inequality.
\item[] $(d)$ $Q\left(u_1^i|u_1^{1:i-1}\right)=\frac{1}{2}$ for $i \in \mathcal{F}_1$.
\item[] $(e)$ $Z(X|Y)^2<H(X|Y)$ \cite{polarsource}.
\item[] $(f)$ Definition of $\mathcal{F}_1$.
\end{itemize}

For the second level, we assume an auxiliary joint distribution $Q'_{U_1^{[N]},U_2^{[N]},Y'^{[N]}}$ resulted from using the encoding rule \eqref{eqn:lossyencoder5} for all $U_1^i$ with $i\in [N]$ at the first partition level, and rules \eqref{eqn:lossyencoder5} and \eqref{eqn:lossyencoder6} at the second level.  Clearly, $Q'_{U_1^{[N]},Y'^{[N]}}= P_{U_1^{[N]},Y'^{[N]}}$ and $Q'_{U_2^{[N]}|U_1^{[N]},Y'^{[N]}} = Q_{U_2^{[N]}|U_1^{[N]},Y'^{[N]}}$. By the triangle inequality,
\begin{eqnarray}
\begin{aligned}
&\mathbb{V}\left(P_{U_1^{[N]},U_2^{[N]},Y'^{[N]}},Q_{U_1^{[N]},U_2^{[N]},Y'^{[N]}}\right) \\
&\leq \mathbb{V}\left(P_{U_1^{[N]},U_2^{[N]},Y'^{[N]}}, Q'_{U_1^{[N]},U_2^{[N]},Y'^{[N]}}\right) \\
&\;\;\;\;\;+ \mathbb{V}\left(Q'_{U_1^{[N]},U_2^{[N]},Y'^{[N]}},Q_{U_1^{[N]},U_2^{[N]},Y'^{[N]}}\right),
\end{aligned}
\end{eqnarray}where the first term on the right hand side (r.h.s.) can be upper bounded by $N\sqrt{\ln2\cdot2^{-N^\beta}}$ using the same method as \eqref{eqn:longTV}, and the second term is equal to $\mathbb{V}\left(P_{U_1^{[N]},Y'{[N]}},Q_{U_1^{[N]},Y'^{[N]}}\right)$.

\begin{figure*}[th]
\begin{eqnarray}\label{eqn:longTV}
\begin{aligned}
&2\mathbb{V}\left(P_{U_1^{[N]},Y'^{[N]}},Q_{U_1^{[N]},Y'^{[N]}}\right) \\
&= \sum_{u_1^{[N]},y^{[N]}} \left|Q(u_1^{[N]},y^{[N]})-P(u_1^{[N]},y^{[N]})\right|\\
&=\sum_{u_1^{[N]},y^{[N]}}\Bigg|\sum_i\left(Q(u_1^i|u_1^{1:i-1},y^{[N]})-P(u_1^i|u_1^{1:i-1},y^{[N]})\right)\\
&\;\;\;\;\;\;\;\;\;\;\;\;\;\;\;\;\;\cdot\left(\prod_{j=1}^{i-1}P(u_1^j|u_1^{1:j-1},y^{[N]})\right)\left(\prod_{j=i+1}^{N}Q(u_1^j|u_1^{1:j-1},y^{[N]})\right)P\Big(y^{[N]}\Big)\Bigg|\\
&\stackrel{(a)}\leq\sum_{i\in\mathcal{F}_1}\sum_{u_1^{[N]},y^{[N]}}\left|Q(u_1^i|u_1^{1:i-1},y^{[N]})-P(u_1^i|u_1^{1:i-1},y^{[N]})\right|\left(\prod_{j=1}^{i-1}P(u_1^j|u_1^{1:j-1},y^{[N]})\right)\\
&\;\;\;\;\;\;\;\;\;\;\;\;\;\;\;\;\;\;\;\;\;\;\cdot\left(\prod_{j=i+1}^{N}Q(u_1^j|u_1^{1:j-1},y^{[N]})\right)P\Big(y^{[N]}\Big)\\
&=\sum_{i\in\mathcal{F}_1}\sum_{u_1^{1:i},y^{[N]}}\left|Q(u_1^i|u_1^{1:i-1},y^{[N]})-P(u_1^i|u_1^{1:i-1},y^{[N]})\right|\left(\prod_{j=1}^{i-1}P(u_1^j|u_1^{1:j-1},y^{[N]})\right)P\Big(y^{[N]}\Big)\\
&=\sum_{i\in\mathcal{F}_1} \sum_{u_1^{1:i-1},y^{[N]}} 2P\left(u_1^{1:i-1},y^{[N]}\right)\mathbb{V}\left(Q_{U_1^i|U_1^{1:i-1}=u_1^{1:i-1},Y'^{[N]}=y^{[N]}},P_{U_1^i|U_1^{1:i-1}=u_1^{1:i-1},Y'^{[N]}=y^{[N]}}\right)\\
&\stackrel{(b)}\leq \sum_{i\in\mathcal{F}_1} \sum_{u_1^{1:i-1},y^{[N]}} P\left(u_1^{1:i-1},y^{[N]}\right) \sqrt{2\ln2 \mathbb{D}_1\left(P_{U_1^i|U_1^{1:i-1}=u_1^{1:i-1},Y'^{[N]}=y^{[N]}}||Q_{U_1^i|U_1^{1:i-1}=u_1^{1:i-1},Y'^{[N]}=y^{[N]}}\right)}\\
&\stackrel{(c)} \leq \sum_{i\in\mathcal{F}_1} \sqrt{2\ln2 \sum_{u_1^{1:i-1},y^{[N]}} P\left(u_1^{1:i-1},y^{[N]}\right) \mathbb{D}_1\left(P_{U_1^i|U_1^{1:i-1}=u_1^{1:i-1},Y'^{[N]}=y^{[N]}}||Q_{U_1^i|U_1^{1:i-1}=u_1^{1:i-1},Y'^{[N]}=y^{[N]}}\right)}\\
&= \sum_{i\in\mathcal{F}_1} \sqrt{2\ln2 \mathbb{D}_1\left(P_{U_1^i}||Q_{U_1^i}|U_1^{1:i-1},Y'^{[N]}\right)}\\
&\stackrel{(d)}= \sum_{i\in\mathcal{F}_1} \sqrt{2\ln2\left(1-H(U_1^i|U_1^{1:i-1},Y'^{[N]})\right)}\\
&\stackrel{(e)}\leq \sum_{i\in\mathcal{F}_1} \sqrt{2\ln2\left(1-Z(U_1^i|U_1^{1:i-1},Y'^{[N]})^2\right)}\\
&\stackrel{(f)}\leq N\sqrt{4\ln2\cdot2^{-N^\beta}}
\end{aligned}
\end{eqnarray}
\hrulefill
\end{figure*}

The proof of the first part of this theorem can be completed by induction. For the second part, when $P_e(\Lambda_r, \tilde{\sigma}_\Delta) \to 0$, the $r+1$-th partition channel is noise free, and its channel capacity $C(\Lambda_r/\Lambda_{r+1}, \tilde{\sigma}_\Delta^2) = 1$. As a result, the $\mathcal{F}_\ell$ is empty for $\ell > r$ by the definition, and the quantization rule \eqref{eqn:lossyencoder5} applies to all indices in $[N]$. Since $Z(X_{\ell}|X_{1:\ell-1},Y')=0$ for $\ell >r$, the distribution $P_{X_{\ell}|X_{1:\ell-1},Y'}$ becomes $\{0,1\}$, and \eqref{eqn:lossyencoder5} becomes deterministic. Then, the distribution $P_{X_{r+1,\cdots}|X_{1:r},Y'}$ turns to extreme, and the rest bits $X_{r+1},\cdots$ can be uniquely determined by rounding $Y'- \mathcal{A}(X_{1:r})$ over $\Lambda_r$, where $\mathcal{A}(X_{1:r})$ denotes the coset of $\Lambda_r$ that labelled by $X_{1:r}$. This statement is similar to the rounding step for the uncoded bits at high levels in the multistage decoding of multilevel coset codes \cite{forney6}.
\end{IEEEproof}

\subsection{Proof of Lemma \ref{lem:EPXY'}}\label{app:EPXY'}
\begin{IEEEproof}
By the i.i.d. property of the vector channel from $X^{[N]}$ to $Y'^{[N]}$, we immediately have $\frac{1}{N}\mathsf{E}_{P}\left[\big\|\alpha Y'^{[N]}-X^{[N]}\big\|^2\right] = \mathsf{E}_{P}\left[\big\|\alpha Y'-X\big\|^2\right]$. Since $Y'-X$ is a Gaussian noise RV that independent of $X$, one can check that $\mathsf{E}_{P}\left[\big\|Y'-X\big\|^2\right] = \Delta$, which can be also expanded as follows.
\begin{eqnarray}
\begin{aligned}
&\mathsf{E}_{P}\left[\big\|Y'-X\big\|^2\right] \\
& = \mathsf{E}_{P}\left[\big\|\alpha Y'-X+(1-\alpha)Y'\big\|^2\right]\\
& = \mathsf{E}_{P}\left[\big\|\alpha Y'-X\big\|^2\right] + (1-\alpha)^2\mathsf{E}_{P}\left[\big\|Y'\big\|^2\right] \\
&\;\;\;\;\;\;\;\;\; + 2(1-\alpha)\mathsf{E}_P\left[(\alpha Y'-X)Y'\right]\\
&= \mathsf{E}_{P}\left[\big\|\alpha Y'-X\big\|^2\right] -(1-\alpha)^2\mathsf{E}_{P}\left[\big\|X\big\|^2\right]+(1-\alpha^2)\Delta,
\end{aligned}
\end{eqnarray}where we use the fact that $\mathsf{E}_{P}\left[\big\|Y'\big\|^2\right] = \mathsf{E}_{P}\left[\big\|X\big\|^2\right]+\Delta$ at the last step.

By \cite[Lem. 6]{cong2}, we have $\left|\mathsf{E}_{P}\left[\|X\|^2\right]-\sigma_r^2\right| \leq \frac{2\pi\epsilon_1}{1-\epsilon_1}\sigma_r^2$. Combining this with the above equality, we obtain
\begin{eqnarray}
\begin{aligned}
&\left|\mathsf{E}_{P}\left[\big\|\alpha Y'-X\big\|^2\right]-\tilde{\sigma}^2_\Delta \right| \\
&= \left|\alpha^2 \Delta +(1-\alpha)^2 \mathsf{E}_{P}\left[\big\|X\big\|^2\right]-\tilde{\sigma}^2_\Delta\right|\\
& \leq \frac{2\pi\epsilon_1}{1-\epsilon_1}(1-\alpha)^2\sigma_r^2 = \frac{2\pi\epsilon_1}{1-\epsilon_1} \frac{\Delta}{\sigma_s^2}\cdot\tilde{\sigma}^2_\Delta.
\end{aligned}
\end{eqnarray}
\end{IEEEproof}

\subsection{Proof of Theorem  \ref{thm:EDistor}}\label{app:EDistor}
\begin{IEEEproof}
\begin{eqnarray}
\begin{aligned}
&\frac{1}{N} \mathsf{E}_{Q}\left[\big\|\alpha Y^{[N]}-X^{[N]}\big\|^2\right] \\
& \leq \frac{1}{N} \sum_{x^{[N]},y^{[N]}} P(x^{[N]},y^{[N]})\big\|\alpha y^{[N]}-x^{[N]}\big\|^2\\
&\;\;\;+ \frac{1}{N} \sum_{x^{[N]},y^{[N]}} |P(\cdot,\cdot)-Q(\cdot,\cdot)| \big\|\alpha y^{[N]}-x^{[N]}\big\|^2\\ \notag
& \leq \frac{1}{N} \sum_{x^{[N]},y^{[N]}} P(x^{[N]},y^{[N]})\big\|\alpha y^{[N]}-x^{[N]}\big\|^2 \\
&\;\;\;+ \frac{2}{N} \mathbb{V}\left(Q_{X^{[N]},Y^{[N]}},P_{X^{[N]},Y'^{[N]}}\right)\cdot N (2^{r-1}V(\Lambda))^2\\
&\leq \tilde{\sigma}^2_\Delta + \frac{2\pi\epsilon_1}{1-\epsilon_1}\frac{\Delta}{\sigma_s^2}\cdot\tilde{\sigma}^2_\Delta + N2^{-N^{\beta'}},
\end{aligned}
\end{eqnarray}where the second inequality is because that the $2^r\Lambda^N$ is a sub-lattice of the constructed polar lattice, and hence $\|\alpha Y -X \| \leq 2^{r-1}V(\Lambda)$ for each dimension; the last inequality is due to Lemma \ref{lem:TVTogether} and Lemma \ref{lem:EPXY'}. Finally, according to Prop. \ref{prop:morelevel} again, we obtain $\epsilon_1 = O(e^{-N})$ by setting $\frac{1}{\eta^2} = O(N)$ and $2^q=\eta 2^r = \sqrt{N}$. The theorem then holds for sufficiently large $N$.
\end{IEEEproof}

\subsection{Proof of Theorem  \ref{thm:VarDistor}}\label{app:VarDistor}
\begin{IEEEproof}
We first show that $\mathsf{E}_{P}\left[\frac{1}{N^2}\big\|\alpha Y'^{[N]}- X^{[N]}\big\|^4\right]$, the abbreviation of the expectation
\[
\mathsf{E}_{P_{X^{[N]},Y'^{[N]}}}\left[\frac{1}{N^2}\big\|\alpha Y'^{[N]}- X^{[N]}\big\|^4\right],
\]is upper-bounded as follows.
\begin{eqnarray}
\mathsf{E}_{P}\left[\frac{1}{N^2}\big\|\alpha Y'^{[N]}- X^{[N]}\big\|^4\right] \leq \Big(1+\frac{2}{N}\Big)\cdot\tilde{\sigma}^4_\Delta\cdot(1+\epsilon_3)^2, \notag
\end{eqnarray} where $\epsilon_3 \triangleq \frac{4(\pi+3)}{3}\frac{\epsilon_2}{1-\epsilon_2}$.

To see this, we expand the norm as
\begin{eqnarray}
\begin{aligned}
&\mathsf{E}_{P}\left[\big\|\alpha Y'^{[N]}- X^{[N]}\big\|^4\right] \\
&= \mathsf{E}_{P} \left[ \sum_{i}(\alpha Y'^i - X^i)^2 \sum_{j}(\alpha Y'^j - X^j)^2 \right]\\ \notag
&= \mathsf{E}_{P} \left[ \sum_{i}(aY'^i-X^i)^4 \right] \\
&\;\;\;+ \mathsf{E}_{P}\left[ \sum_{i}\sum_{j \neq i} (\alpha Y'^i - X^i)^2 (\alpha Y'^j - X^j)^2 \right]\\
&= N\cdot \mathsf{E}_{P} \left[ \|aY'-X\|^4 \right] + N(N-1)\cdot \mathsf{E}^2_{P} \left[ \|aY'-X\|^2 \right].
\end{aligned}
\end{eqnarray}
For the first term on the r.h.s.,
\begin{eqnarray}
\begin{aligned}
\mathsf{E}_{P} \left[ \|aY'-X\|^4 \right] &= (1-\alpha)^4 \mathsf{E}_{P}\left[ \|X\|^4 \right] \\ \notag
&\;\;\; + 6\alpha^2 (1-\alpha)^2\Delta \mathsf{E}_{P}\left[\|X\|^2\right] + 3\alpha^4\Delta^2.
\end{aligned}
\end{eqnarray}
One may check that when $X$ is a continuous Gaussian RV with zero mean and variance $\sigma_r^2$, $\mathsf{E}_{P} \left[ \|aY'-X\|^4 \right] = 3\tilde{\sigma}^4_\Delta$, since $\alpha Y' - X $ is also a continuous Gaussian RV with zero mean and variance $\tilde{\sigma}^2_\Delta$ in this case. When $X$ is sampled from $D_{\Lambda,\sigma_r}$ instead, by Lemma \ref{lem:4thGap} and \cite[Lem. 6]{cong2},
\begin{eqnarray}
\begin{aligned}
&\mathsf{E}_{P} \left[ \|aY'-X\|^4 \right] \\
&\leq (1-\alpha)^4 \cdot 3\sigma_r^4 \left(1+\frac{4(\pi+3)}{3}\frac{\epsilon_2}{1-\epsilon_2}\right) \\
&\;\;\;+ 6\alpha^2 (1-\alpha)^2\Delta \cdot \sigma_r^2 \left(1+\frac{2\pi\epsilon_1}{1-\epsilon_1}\right) + 3\alpha^4\Delta^2\\
& \leq 3\tilde{\sigma}^4_\Delta \left(1+\frac{4(\pi+3)}{3}\frac{\epsilon_2}{1-\epsilon_2}\right).
\end{aligned}
\end{eqnarray}

Using the upper-bound in Lemma \ref{lem:EPXY'} for $\mathsf{E}_{P} \left[ \|aY'-X\|^2 \right]$, we have
\begin{eqnarray}
\begin{aligned}
\mathsf{E}_{P}^2 \left[ \|aY'-X\|^2 \right] &\leq \left(1+\frac{2\pi\epsilon_1}{1-\epsilon_1}\frac{\Delta}{\sigma_s^2}\right)^2\tilde{\sigma}^4_\Delta \\
&\leq \left(1+\frac{2\pi\epsilon_1}{1-\epsilon_1}\right)^2\tilde{\sigma}^4_\Delta.
\end{aligned}
\end{eqnarray}
Letting $\epsilon_3 = \frac{4(\pi+3)}{3}\frac{\epsilon_2}{1-\epsilon_2}$, and recalling that $\epsilon_1 \leq \epsilon_2$,
\begin{eqnarray}
\begin{aligned}
&\mathsf{E}_{P}\left[\big\|\alpha Y'^{[N]}- X^{[N]}\big\|^4\right] \\
&\leq N\cdot 3 \tilde{\sigma}^4_\Delta (1 + \epsilon_3) + N(N-1)\cdot \left(1+\frac{2\pi\epsilon_1}{1-\epsilon_1}\right)^2\tilde{\sigma}^4_\Delta \\
&\leq N(N+2)\cdot \tilde{\sigma}^4_\Delta \cdot (1+\epsilon_3)^2.
\end{aligned}
\end{eqnarray}

Next, by using the similar idea of the proof in Theorem \ref{thm:EDistor},
\begin{eqnarray}
\begin{aligned}
&\mathsf{Var}_Q\left[\frac{1}{N}\big\|\alpha Y^{[N]}-X^{[N]}\big\|^2\right] \\
&= \mathsf{E}_Q \left[\frac{1}{N^2}\big\|\alpha Y^{[N]}-X^{[N]}\big\|^4\right] - \mathsf{E}^2_Q \left[\frac{1}{N}\big\|\alpha Y^{[N]}-X^{[N]}\big\|^2\right] \\
& \leq \mathsf{E}_P \left[\frac{1}{N^2}\big\|\alpha Y^{[N]}-X^{[N]}\big\|^4\right] \\
&\;\;\;+ \frac{1}{N^2} \sum_{x^{[N]},y^{[N]}} |P(\cdot,\cdot)-Q(\cdot,\cdot)| \big\|\alpha y^{[N]}-x^{[N]}\big\|^4 - \tilde{\sigma}^4_\Delta\\ \notag
&\leq \Big(1+\frac{2}{N}\Big)\cdot\tilde{\sigma}^4_\Delta\cdot(1+\epsilon_3)^2 \\
&\;\;\;+ \frac{2}{N^2} \mathbb{V}(Q,P)\cdot \Big(N \big(2^{r-1}V(\Lambda)\big)^2\Big)^2 - \tilde{\sigma}^4_\Delta \\
&\leq \frac{6}{N} \tilde{\sigma}^4_\Delta \cdot \epsilon_3 + 2^{4q-2} \cdot 2^{-N^{\beta'}}\\
&\leq N^2 \cdot 2^{-N^{\beta'}}.
\end{aligned}
\end{eqnarray}

\end{IEEEproof}

\subsection{Proof of Lemma \ref{lem:4thGap}}\label{app:4thGap}
\begin{IEEEproof}
Our proof follows the steps of that in \cite[Lem. 4.2]{MicciancioRegev04}. For convenience, we first consider $s=\sqrt{2\pi}\sigma_r =1$ and then scale the result back. It can be checked that $\widehat{g_4}(y) = (y^4 - \frac{3}{\pi}y^2 + \frac{3}{4\pi^2}) \widehat{\rho_c}(y)$, where $\widehat{\rho_c}(y)$ is defined as $\widehat{\rho_c}(y)=\rho(y)e^{-2\pi i <y,c>}$ for the standard Gaussian function $\rho_c(y)$ with factor $s=1$, and $\widehat{g_4}(y)$ is the Fourier transform of the function $g_4(x) = (x-c)^4\cdot \rho_c(x)$. Then, according to \cite[eq. (7)]{MicciancioRegev04}, we have
\begin{eqnarray}
\begin{aligned}
&\bigg|\mathsf{E}\left[(x-c)^4\right]-\frac{3}{4\pi^2}\bigg| \\
&= \bigg|\frac{\widehat{g_4}(y)}{\widehat{\rho_c}(y)} - \frac{3}{4\pi^2} \bigg| \\
& = \frac{\big|\sum_{y\in\Lambda^*}(y^4-\frac{3}{\pi}y^2) \widehat{\rho_c}(y)\big|}{\widehat{\rho_c}(\Lambda^*)}\\
&\leq \frac{\sum_{y\in\Lambda^*}y^4 \rho(y)}{1-\rho(\Lambda^*\setminus\{0\})} + \frac{3}{\pi}\cdot\frac{\sum_{y\in\Lambda^*}y^2 \rho(y)}{1-\rho(\Lambda^*\setminus\{0\})},
\end{aligned}
\end{eqnarray}where the numerator of the first term on the r.h.s. can be upper bounded as the following, using the bound $y^2 \leq e^{y^2/e}$.
\begin{eqnarray}
\begin{aligned}
\sum_{y\in\Lambda^*}y^4 \rho(y) &\leq \sum_{y\in\Lambda^*} e^{2y^2/e} \cdot e^{-\pi y^2} \\
&= \sum_{y\in\Lambda^*} e^{-(\pi-2/e)y^2} = \epsilon_{\Lambda}\left(\sigma_r/\sqrt{\frac{\pi}{\pi-2/e}}\right).
\end{aligned}
\end{eqnarray}
Similarly, for the second term, we have
\begin{eqnarray}
\begin{aligned}
\sum_{y\in\Lambda^*}y^2 \rho(y) &\leq \sum_{y\in\Lambda^*} e^{y^2/e} \cdot e^{-\pi y^2} \\
&= \sum_{y\in\Lambda^*} e^{-(\pi-1/e)y^2} = \epsilon_{\Lambda}\left(\sigma_r/\sqrt{\frac{\pi}{\pi-1/e}}\right).
\end{aligned}
\end{eqnarray}
One can check that $\epsilon_{\Lambda}\left(\sigma_r/\sqrt{\frac{\pi}{\pi-1/e}}\right) < \epsilon_{\Lambda}\left(\sigma_r/\sqrt{\frac{\pi}{\pi-2/e}}\right) = \epsilon_2$. By \cite[eq. (8)]{MicciancioRegev04}, $1-\rho(\Lambda^*\setminus\{0\})$ is lower bounded by $1-\epsilon_2$ in the same fashion. Therefore, we arrive at
\begin{eqnarray}
\bigg|\mathsf{E}\left[(x-c)^4\right]-\frac{3}{4\pi^2}\bigg| \leq \frac{\epsilon_2}{1-\epsilon_2}\left(1+\frac{3}{\pi}\right).
\end{eqnarray}
By scaling $s$ back to $\sqrt{2\pi}\sigma_r$, we complete the proof.
\end{IEEEproof}

\subsection{Proof of Lemma \ref{lem:SecMBound}}\label{app:SecMBound}
\begin{IEEEproof}
Let $\frac{1}{N}\big\|\alpha Y^{[N]}-X^{[N]}\big\|^2$ be abbreviated as $*$ for convenience. By Chebyshev's inequality, for any $\delta >0$,
\begin{eqnarray}\label{eqn:mainIneq}
\mathsf{Prob.}\left(\left|*-\mathsf{E}_{Q}[*]\right| \geq \delta \right) \leq  \frac{\mathsf{Var}_Q\left[*\right]}{\delta^2}.
\end{eqnarray}
Therefore, $\mathsf{Prob.}\left(\big\|\alpha Y^{[N]}-X^{[N]}\big\|^2 \leq N\cdot\mathsf{E}_{Q}[*] + N \cdot \delta \right) \geq 1- \frac{\mathsf{Var}_Q[*]}{\delta^2}$. We notice that the above inequality holds for any Gaussian source $Y\sim f_{\sigma_s}$. We can fix $\tilde{\sigma}_\Delta^2$ and set $\Delta = \frac{\sigma_s^2-\sqrt{\sigma_s^4-4\sigma_s^2\tilde{\sigma}^2_\Delta}}{2}$ for a given large $\sigma_s^2$. The compression rate is $\frac{1}{2}\log \left(\frac{\sigma_s^2}{\Delta}\right)=\frac{1}{2}\log \left(\frac{2}{1-\sqrt{1-{4\tilde{\sigma}_\Delta^2}/{\sigma_s^2}}}\right) \approx \frac{1}{2}\log\frac{\sigma_s^2}{\tilde{\sigma}_\Delta^2}$, which corresponds to the high-resolution quantization region of the Gaussian source.

Let $\mathcal{P}(\Lambda_Q)$ denote the quantization region of $\Lambda_Q$ w.r.t. the rule given in \eqref{eqn:lossyencoder5}. By the definition of $\mathcal{I}_\ell$, the probability $P_{U_{\ell}^{i}|U_{\ell}^{1:i-1},X_{1:\ell-1}^{[N]},Y'^{[N]}}$ turns to the extreme distribution $\{0,1\}$ as $N$ increases for indices with $Z\left(U_{\ell}^{i}|U_{\ell}^{1:i-1},X_{1:\ell-1}^{[N]}, Y'^{[N]}\right)\leq 2^{-N^{\beta}}$, and the randomness of $\mathcal{P}(\Lambda_Q)$ is caused by those with $Z\left(U_{\ell}^{i}|U_{\ell}^{1:i-1},X_{1:\ell-1}^{[N]}, Y'^{[N]}\right) \in \left(2^{-N^{\beta}}, 1-2^{-N^{\beta}}\right)$, whose proportion is vanishing as $N$ increases. By the high-resolution quantization theory \cite{Bennett1948,GrayQZnoise90}, the conditional distribution of the vector $\alpha Y^{[N]}$, given the $\alpha Y^{[N]}$ falls into $\mathcal{P}(\Lambda_Q)$, is roughly uniform in $\mathcal{P}(\Lambda_Q)$, and the quantization noise is commonly modeled as an independent uniform noise to the source.

Our strategy is to use \eqref{eqn:mainIneq} to give an upper bound on the expectation $\mathsf{E}_Q\left[\sigma^2(\mathcal{P}(\Lambda_Q))\right]$ of the second moment $\sigma^2(\mathcal{P}(\Lambda_Q))$, where we similarly define $\sigma^2(\mathcal{P}(\Lambda_Q)) \triangleq \frac{1}{NV(\mathcal{P}(\Lambda_Q)}{\int_{\mathcal{P}(\Lambda_Q)}\|\mathbf{u}\|^2 d\mathbf{u}}$. Since $V(\mathcal{P}(\Lambda_Q)) = V(\Lambda_Q)$, we also have a trivial upper bound $\sigma^2(\Lambda_Q) \leq \mathsf{E}_Q\left[\sigma^2(\mathcal{P}(\Lambda_Q))\right]$ by the fact that the Voronoi region minimizes the second moment for all fundamental regions \cite[Lem. 4.3.1]{BK:Zamir}. We note that the difference between $\mathcal{P}(\Lambda_Q)$ and $\mathcal{V}(\Lambda_Q)$ is due to the multi-level decoding of polar lattices. However, the performance of the multi-level lattice decoding converges to that of the optimal lattice decoding as the channel polarization effect becomes sufficient when $N \to \infty$, which means the upper bound is tight. A demonstration of $\mathcal{V}(\Lambda_Q)$ and $\mathcal{P}(\Lambda_Q)$ in the two-dimensional case is depicted in Fig.\ref{fig:QZregion}.

\begin{figure}[ht]
    \centering
    \includegraphics[width=6cm]{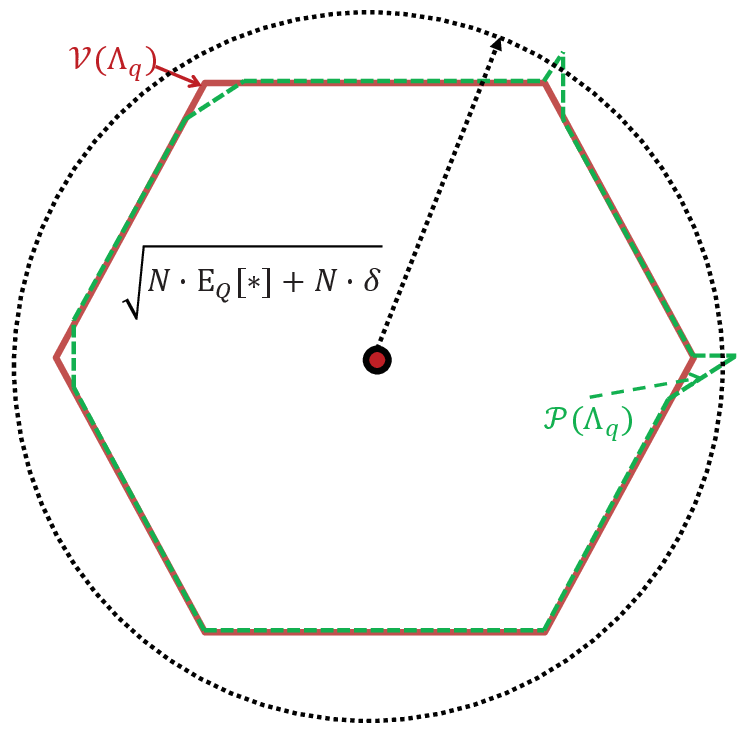}
    \caption{A demonstration of $\mathcal{V}(\Lambda_Q)$, $\mathcal{P}(\Lambda_Q)$ and the sphere $\mathcal{B}_{\sqrt{N\cdot\mathsf{E}_{Q}[*] + N \cdot \delta}}$ in the two-dimensional case.}
    \label{fig:QZregion}
\end{figure}

Let $E$ denote the event that $\big\|\alpha Y^{[N]}-X^{[N]}\big\|^2 \leq N\cdot\mathsf{E}_{Q}[*] + N \cdot \delta$ and $\bar{E}$ denote its complement event. We have
\begin{eqnarray}
\begin{aligned}
\mathsf{E}_Q\left[\sigma^2(\mathcal{P}(\Lambda_Q))\right]  &= \mathsf{Prob.}(E) \cdot \sigma^2(\mathcal{P}(\Lambda_Q)|E) \\
&\;\;\;+ \mathsf{Prob.}(\bar{E}) \cdot \sigma^2(\mathcal{P}(\Lambda_Q)|\bar{E}).
\end{aligned}
\end{eqnarray}where $\sigma^2(\mathcal{P}(\Lambda_Q)|E)$ and $\sigma^2(\mathcal{P}(\Lambda_Q)|\bar{E})$ denote the second moment of $\mathcal{P}(\Lambda_Q)$ under the condition of $E$ and $\bar{E}$, respectively.

For the first term, using the reverse iso-perimetric inequalities, $\sigma^2(\mathcal{P}(\Lambda_Q)|E) \leq \sigma^2\left(\mathcal{B}_{\sqrt{N\mathsf{E}_{Q}[*] + N\delta}}\right)$, where $\mathcal{B}_r$ denotes the $N$-dimensional sphere with radius $r$. For the second term, since $\Lambda'^N \subseteq \Lambda_Q \subseteq \Lambda^N$, we have $\|\mathbf{u}\|^2 \leq N\left(\frac{V(\Lambda')}{2}\right)^2$, and hence $\frac{\int_{\mathcal{P}(\Lambda_Q)}\|\mathbf{u}\|^2 d\mathbf{u}}{NV(\Lambda_Q)} \leq \frac{N\left(2^{r-1}V(\Lambda)\right)^2 V(\Lambda_Q)}{N V(\Lambda_Q)}$. Putting these two upper bounds together, we obtain
\begin{eqnarray}
\begin{aligned}
\sigma^2&(\Lambda_Q) \\
&\leq \mathsf{Prob.}(E) \cdot \sigma^2(\mathcal{P}(\Lambda_Q)|E) + \mathsf{Prob.}(\bar{E}) \cdot \sigma^2(\mathcal{P}(\Lambda_Q)|\bar{E})\\
&\leq \left(1- \frac{\mathsf{Var}_Q}{\delta^2}\right) \sigma^2\left(\mathcal{B}_{\sqrt{N\mathsf{E}_{Q}[*] + N \delta}}\right) + \frac{\mathsf{Var}_Q}{\delta^2}(2^{r-1}V(\Lambda))^2\\
& \leq \frac{N\cdot\mathsf{E}_{Q}[*] + N \cdot \delta}{N+2} + \frac{\mathsf{Var}_Q}{\delta^2}(2^{r-1}\eta)^2\\ \notag
&\leq \frac{N}{N+2} (\tilde{\sigma}^2_\Delta+\delta) + \frac{2N^2}{N+2} 2^{-N^{\beta'}} + \frac{N^2 2^{-N^{\beta'}}}{\delta^2}\cdot N\\
&\leq \frac{N}{N+2}\tilde{\sigma}^2_\Delta + \frac{2N^2}{N+2} 2^{-N^{\beta'}} +3N\cdot2^{-\frac{1}{3}N^{\beta'}} \\
&\leq \frac{N}{N+2}\left(\tilde{\sigma}^2_\Delta + 2^{-N^{\beta''}}\right),
\end{aligned}
\end{eqnarray}where we use the fact $\sigma^2(\mathcal{B}_r) = \frac{r^2}{N+2}$ in the third inequality. The fourth inequality holds because of Theorem \ref{thm:EDistor} and Theorem \ref{thm:VarDistor}. For the fifth inequality, since this bound holds for any $0< \delta <1$, we may choose $\delta = \sqrt[3]{N^2(N+2)2^{-N^{\beta'}}}$ such that $\frac{2N}{N+2}\delta + \frac{N^2 2^{-N^{\beta'}}}{\delta^2}N = 3N\cdot\sqrt[3]{\frac{N^2}{(N+2)^2}2^{-N^{\beta'}}}$ by the AM-GM inequality. The last inequality holds by letting $2^{-N^{\beta''}} = 6(N+2)\cdot2^{-\frac{1}{3}N^{\beta'}}$ for some $\beta'' < \beta'$ and sufficiently large $N$.
\end{IEEEproof}

\subsection{Proof of Lemma \ref{lem:VNRbound}}\label{app:VNRbound}

\begin{IEEEproof}
Since $\Lambda_Q$ is constructed from the partition chain in Fig. \ref{fig:1Dpartition} according to the construction D method, we have $\Lambda'^N \subseteq \Lambda_Q \subseteq \Lambda^N$. Let $R_{\mathcal{C}} = \sum_{\ell=1}^r R_\ell = \frac{1}{N} \sum_{\ell=1}^r |\mathcal{I}_\ell|$ denote the total coding rate of the component polar codes according to \eqref{eqn:BitAssign}. Then, $V(\Lambda_Q) = 2^{-N R_{\mathcal{C}}} V(\Lambda')^N$ as proven in \cite[Sect. V-A]{forney6}. The logarithmic VNR of $\Lambda_Q$ is
\begin{eqnarray}
\begin{aligned}
\log\left(\frac{\gamma_{\Lambda_Q}(\tilde{\sigma}_\Delta)}{2\pi e}\right) &= \log \frac{V(\Lambda_Q)^{\frac{2}{n_pN}}}{2\pi e \tilde{\sigma}^2_\Delta} \\
&= \log \frac{2^{-2R_{\mathcal{C}}}V(\Lambda')^2}{2\pi e \tilde{\sigma}^2_\Delta} \\
& = -2R_{\mathcal{C}} + 2 \log V(\Lambda') - \log(2\pi e \tilde{\sigma}^2_\Delta),
\end{aligned}
\end{eqnarray}where we use the condition $n_p = 1$. Define
\begin{eqnarray}
\begin{cases} \epsilon_{a}=C(\Lambda,\tilde{\sigma}^2_\Delta) \\
\epsilon_{b}=h(\tilde{\sigma}^2_\Delta)-h(\Lambda',\tilde{\sigma}^2_\Delta) \\\notag
\epsilon_{c}=R_{\mathcal{C}}-C(\Lambda/\Lambda', \tilde{\sigma}^2_\Delta)=\sum_{\ell=1}^{r}R_{\ell}-{C(\Lambda_{\ell-1}/\Lambda_{\ell}, \tilde{\sigma}^2_\Delta)},
\end{cases}
\label{eqn:epsilons}
\end{eqnarray}where $h(\tilde{\sigma}^2_\Delta)$ and $h(\Lambda',\tilde{\sigma}^2_\Delta)$ are the differential entropies of the continuous Gaussian noise and the $\Lambda'$-aliased Gaussian noise defined in Sect. \ref{sec:background}, respectively. $C(\Lambda,\tilde{\sigma}^2_\Delta)$ and $C(\Lambda_{\ell-1}/\Lambda_{\ell}, \tilde{\sigma}^2_\Delta)$ are the capacities of the mod-$\Lambda$ channel and the $\Lambda_{\ell-1}/\Lambda_{\ell}$ channel defined in \cite{forney6}. We see that $\epsilon_a \geq 0$ is the capacity of the top mod-$\Lambda$ channel, $\epsilon_b$ is the entropy loss of the Gaussian noise after the mod-$\Lambda'$ operation at the bottom level, and $\epsilon_c$ is the total overhead of the compression rate of the component polar codes for the $r$ levels.
Then, we have
\begin{eqnarray}
\log\left(\frac{\gamma_{\Lambda_Q}(\tilde{\sigma}_\Delta)}{2\pi e}\right) = 2 (\epsilon_a - \epsilon_b - \epsilon_c) \geq  -2(\epsilon_b + \epsilon_c).
\end{eqnarray}
\end{IEEEproof}

\subsection{Proof of Theorem \ref{thm:QZgood}}\label{app:QZgood}
\begin{IEEEproof}
From Lemma \ref{lem:VNRbound}, we have
\begin{eqnarray}
V(\Lambda_Q)^{\frac{2}{N}} \geq 2 \pi e \tilde{\sigma}^2_\Delta \cdot 2^{-2(\epsilon_b +\epsilon_c)}.
\end{eqnarray}
Combining this with Lemma \ref{lem:SecMBound}, and by the definition of NSM,
\begin{eqnarray}\label{eqn:NSMeq1}
\begin{aligned}
G(\Lambda_Q) &\leq \frac{\frac{N}{N+2}{\left(\tilde{\sigma}^2_\Delta+2^{-N^{\beta''}}\right)}}{2 \pi e \tilde{\sigma}^2_\Delta \cdot 2^{-2(\epsilon_b +\epsilon_c)}} \\
&= \frac{N}{N+2} 2^{2(\epsilon_b +\epsilon_c)} \frac{1}{2\pi e} \left(1+ \frac{2^{-N^{\beta''}}}{\tilde{\sigma}^2_\Delta}\right).
\end{aligned}
\end{eqnarray}

Let $f_{\sigma}$ and $f_{\sigma, \Lambda'}$ denote the standard Gaussian distribution and the $\Lambda'$-aliased Gaussian distribution, respectively. We see that $\epsilon_b$ is resulted from the difference between the differential entropies of $f_{\sigma}$ and $f_{\sigma, \Lambda'}$. The distribution $f_{\sigma, \Lambda'}$ can be viewed as the result of modifying $f_{\sigma}$ by transporting its density, which lies outside of $\mathcal{V}(\Lambda')$, into $\mathcal{V}(\Lambda')$ according to the mod-$\Lambda'$ operation. Therefore, since $\Lambda' = 2^q \mathbb{Z}$ in Fig. \ref{fig:1Dpartition},
\begin{eqnarray}
\begin{aligned}
\mathbb{V}(f_{\sigma}, f_{\sigma, \Lambda'}) &\leq \int_{-\infty}^{-2^{q-1}} f_{\sigma}(x) dx + \int_{2^{q-1}}^{\infty} f_{\sigma}(x) dx\\
&\leq 2 \frac{1}{\sqrt{2\pi}\sigma}\int_{2^{q-1}}^\infty \exp\left(-\frac{x^2}{2\sigma^2}\right)dx\\
&= 2 \cdot \mathsf{Q}\left(\frac{2^{q-1}}{\sigma}\right) \\
&\leq 2 \cdot \exp\left(-\frac{2^{2q}}{8\sigma^2}\right),
\end{aligned}
\end{eqnarray}where $\mathsf{Q}(x)$ is the Q-function of a standard normal distribution, and we use $\mathsf{Q}(x)\leq \exp(-\frac{x^2}{2})$ in the last inequality.

Recalling that $2^q=\sqrt{N}$ and using \cite[Thm. 1]{HamidTVConti17},
\begin{eqnarray}
\begin{aligned}
\epsilon_b &\leq 2c_1 \cdot\mathbb{V}(f_{\tilde{\sigma}_\Delta}, f_{\tilde{\sigma}_\Delta, \Lambda'}) \\
&\;\;\;+ 2c_2 \cdot\mathbb{V}(f_{\tilde{\sigma}_\Delta}, f_{\tilde{\sigma}_\Delta, \Lambda'}) \ln \frac{1}{\mathbb{V}(f_{\tilde{\sigma}_\Delta}, f_{\tilde{\sigma}_\Delta, \Lambda'})},
\end{aligned}
\end{eqnarray}where $c_1$ and $c_2$ are two positive constants independent of $N$, and $\ln$ is the natural logarithm. One can set $n=1$, $\alpha=2$, $v=2\tilde{\sigma}^2_\Delta$, and $m = \frac{2}{\sqrt{2\pi\tilde{\sigma}^2_\Delta}}$ to get $c_1 $ and $c_2$ as in \cite[eq.(14)]{HamidTVConti17} and \cite[eq.(15)]{HamidTVConti17}, respectively. For sufficiently large $N$, the second term dominates on the r.h.s. of the above inequality. Therefore, $\epsilon_b \leq c_2\cdot \frac{N}{\tilde{\sigma}^2_\Delta}\exp\left(-\frac{N}{8\tilde{\sigma}^2_\Delta}\right)$.

Now we turn to the bounds on the finite length scaling of polar codes to derive an upper bound for $\epsilon_c$. For each partition level, we use the same idea of the proof of \cite[Thm. 2]{GoldinScal14}. To be brief, we follow the notations in \cite{GoldinScal14}. We can derive a lower bound on $\mathcal{D}_N(R)$ from the other direction as in the inequality above \cite[eq. (39)]{GoldinScal14}.
\begin{eqnarray}
\begin{aligned}
\mathcal{D}_N(R) &= D_N - D + D - D(R) \\
&= D_N - D + \Delta_\ell \left| \frac{D(R)-D}{R-C(\Lambda_{\ell-1}/\Lambda_{\ell}, \tilde{\sigma}^2_\Delta)} \right|  \\
&\geq  D_N - D + \Delta_\ell |D'(R)|,
\end{aligned}
\end{eqnarray}where $\mathcal{D}_N(R) \triangleq D_N - D(R)$ denotes the extra distortion introduced by using polar codes, $\Delta_\ell \triangleq R - C(\Lambda_{\ell-1}/\Lambda_{\ell},\tilde{\sigma}^2_\Delta)$ for each partition level, and the last inequality is due to the convexity of the distortion-rate function $D(R)$.
Then, using the result of \cite[Thm. 2]{GoldinScal14},
\begin{eqnarray}
N = \frac{\beta_\ell}{\mathcal{D}_0^{\mu}} \leq \frac{\beta_\ell}{\mathcal{D}_N(R)^{\mu}} \leq \frac{\bar{\beta}_\ell}{\Delta_\ell^\mu},
\end{eqnarray}where $\mathcal{D}^0$ is the upper bound of $\mathcal{D}_N(R)$, and $\bar{\beta}_\ell =\frac{\beta_\ell}{|D'(R)|^\mu}$ is a constant that depends only on $R_\ell$, $d_{max}$ and the distortion measure function. Therefore, $\Delta_\ell \leq (\frac{\bar{\beta}_\ell}{N})^{\frac{1}{\mu}}$ for level $1 \leq \ell \leq r$. By choosing $\bar{\beta}_{max} = \max\{\bar{\beta}^{1/\mu}_1,...,\bar{\beta}^{1/\mu}_r\}$, we have
\begin{eqnarray}
\epsilon_c = \sum^r_{\ell=1} \Delta_{\ell} \leq r \cdot\frac{\bar{\beta}_{max}}{N^{\frac{1}{\mu}}}.
\end{eqnarray}

By setting that $\frac{1}{\eta^2} = O(N)$ and $r =O(\log N)$ as in Prop. \ref{prop:morelevel}, we have $\epsilon_b = O(Ne^{-N})$, which is dominated by $\epsilon_c = O\left(\frac{\log N }{N ^\frac{1}{\mu}}\right)$. Therefore, by plugging the forms of $\epsilon_b$ and $\epsilon_c$ into \eqref{eqn:NSMeq1},
\begin{eqnarray}
\lim_{N\to \infty} G(\Lambda_Q) \leq \frac{1}{2\pi e},
\end{eqnarray}and the equality holds due to the trivial lower bound $G(\Lambda_Q) \geq \frac{1}{2\pi e}$. Moreover,
\begin{eqnarray}
\begin{aligned}
\lim_{N\to \infty}\log(G(\Lambda_Q)\cdot 2\pi e) &\leq 2\epsilon_b+ 2\epsilon_c + \frac{2^{-N^{\beta''}}}{\tilde{\sigma}^2_\Delta}\log e \\
&= O\left(\frac{\log N}{N^\frac{1}{\mu}}\right),
\end{aligned}
\end{eqnarray}where we use the inequality $\log(1+x) \leq x\cdot\log (e)$.
\end{IEEEproof}

\subsection{Proof of Theorem \ref{thm:fixF}}\label{app:fixF}
\begin{IEEEproof}
Let $\frac{1}{N}\big\|\alpha Y^{[N]}-X^{[N]}\big\|^2$ be abbreviated as $*$ for convenience. By Chebyshev's inequality, for any $\delta_1 >0$,
\begin{eqnarray}
\mathsf{Prob.}\left(\left|\mathsf{E}_{Q|U^{\mathcal{F}_{\Lambda_Q}}}\left[*\right]-\mathsf{E}_{Q}\left[*\right]\right| \geq \delta_1 \right) \\
 &\hspace{-5em}  \leq  \frac{\mathsf{Var}_{U^{\mathcal{F}_{\Lambda_Q}}}\left[\mathsf{E}_{Q|U^{\mathcal{F}_{\Lambda_Q}}}[*]\right]}{\delta_1^2},
\end{eqnarray}where $\mathsf{Var}_{U^{\mathcal{F}_{\Lambda_Q}}}\left[\mathsf{E}_{Q|U^{\mathcal{F}_{\Lambda_Q}}}[*]\right]$ is the variance of $\mathsf{E}_{Q|U^{\mathcal{F}_{\Lambda_Q}}}\left[*\right]$ w.r.t. $U^{\mathcal{F}_{\Lambda_Q}}$.

Using the law of total variance, we have
\begin{eqnarray}\label{eqn:TotalVar}
\begin{aligned}
\mathsf{Var}_Q[*] &= \mathsf{E}_{U^{\mathcal{F}_{\Lambda_Q}}}\left[\mathsf{Var}_{Q|U^{\mathcal{F}_{\Lambda_Q}}}[*]\right]  \\
&\;\;\;\;\;\;\;\;\;\;\;\;\;\;\;\;\;\; + \mathsf{Var}_{U^{\mathcal{F}_{\Lambda_Q}}}\left[\mathsf{E}_{Q|U^{\mathcal{F}_{\Lambda_Q}}}[*]\right], 
\end{aligned}
\end{eqnarray}which means $\mathsf{Var}_{U^{\mathcal{F}_{\Lambda_Q}}}\left[\mathsf{E}_{Q|U^{\mathcal{F}_{\Lambda_Q}}}[*]\right] \leq \mathsf{Var}_Q[*]$ by the  non-negativity of variance.

Therefore, by setting $\delta_1 = \sqrt[3]{N^22^{-N^{\beta'}}}$ and invoking Theorem \ref{thm:VarDistor}, the proof is completed.
\end{IEEEproof}

\end{document}